\documentclass[utf8]{aa}
\usepackage{amsmath}
\usepackage{amsfonts}   
\usepackage{amssymb}
\usepackage{siunitx}
\usepackage{commath}
\usepackage{color}
\usepackage{pdflscape}
\usepackage{natbib}
\usepackage{graphicx}
\usepackage{verbatim}
\bibpunct{(}{)}{;}{a}{}{,}
\begin{document}

   \title{Spatially resolved star formation relation in two HI-rich galaxies with central post-starburst signature}
   \titlerunning{Spatially resolved star formation relation in two HI-rich galaxies}


   \author{Anne~Klitsch
          \inst{1,2}
          \and
          Martin~A.~Zwaan\inst{1}
          \and
          Harald~Kuntschner\inst{1}
          \and
          Warrick~J.~Couch\inst{3}
          \and
          Michael~B.~Pracy\inst{4}          
          \and
          Matt Owers\inst{5,6}
          }
    \authorrunning{A.~Klitsch et al.}

   \institute{European Southern Observatory, Karl-Schwarzschild-Str. 2, 85748 Garching bei München, Germany\\
              \email{aklitsch@eso.org}
         \and
             Ludwig-Maximilians-Universität München, Faculty of Physics, Schellingstraße 4, 80799 München, Germany
         \and
                Australian Astronomical Observatory, PO Box 915, North Ryde, NSW 1670, Australia
                \and 
                        Sydney Institute for Astronomy, School of Physics, University of Sydney, NSW 2006, Australia
         \and
         Department of Physics and Astronomy, Macquarie University, NSW,
2109, Australia
		\and
		The Australian Astronomical Observatory, PO Box 915, North Ryde,
NSW, 1670, Australia
             }

   \date{Draft of \today}


  \abstract
   {E+A galaxies are post-starburst systems that are identified based on their optical spectra. These galaxies contain a substantial
young A-type stellar component but display no emission lines,
which indicates only little ongoing star formation (SF). HI 21
cm line emission is found in approximately half of the nearby E+A galaxies, indicating that they contain a reservoir of gas that could fuel active SF. 
   }
   {We study the distribution and kinematics of atomic and molecular gas in two HI-rich galaxies, which show a typical E+A spectrum at the centre and SF at larger radii. From our results we aim to infer whether the SF activity of these galaxies is consistent with the activity seen in disc galaxies, or if it indicates a transition towards another evolutionary phase. 
   }
   {We present newly obtained high spatial resolution radio interferometric observations of the HI~21 cm emission line using the Karl Jansky Very Large Array (VLA) and of the CO(1-0) emission line using the Atacama Large Millimeter/submillimeter Array (ALMA). We combine these data sets to predict the star formation rate (SFR) using a pressure-based SF relation and show that it does not correlate well with the SFR derived from H$\alpha$ on sub-kpc scales. We apply a recently developed statistical model for the small-scale behaviour of the SF relation to predict and interpret the observed scatter.
  }
   {We find regularly rotating HI gas that is smoothly
distributed across the entire disc. The CO(1-0) emission line is not detected for either of the two galaxies. The derived upper limit on the CO mass implies a molecular gas depletion time of $t_{\rm depl} \lesssim 20 \; {\rm Myr}$. However, because of the low metallicity, the CO-to-H2 conversion factor is highly uncertain. In the relations between the H$\alpha$-based SFR and the HI mass, we observe a substantial scatter we demonstrate results from small-number statistics of independent SF regions on sub-kpc scales. 
   }
   {We confirm the HI-richness of ESO534-G001 and 2dFRS~S833Z022, and find that the scale dependence of the atomic SF relation in these galaxies is consistent with the predicted increase in the scatter towards small size scales. This is due to the incomplete sampling of independent HI clouds and SF regions. This finding adds to the existing literature, which has reported a scale dependence of the \textit{\textup{molecular}} SF relation, showing that the atomic and molecular phases are both susceptible to the evolutionary cycling of individual regions. This suggests that the atomic gas reservoirs host substantial substructure, which should be observable with future high-resolution observations.
   
   }

   \keywords{Stars: formation--
                                Galaxies: evolution--
                                Galaxies: ISM
               }

   \maketitle

%

\section{Introduction}

        \begin{table*}
\caption{\label{TabGalProp}Summary of galaxy properties}
\centering
\begin{tabular}{llccccccccc}
\hline\hline
&Name & RA & $V_{\si{sys}}$ & $S_{\si{int}}$ & $S_{\si{peak}}$ & $M_{\si{HI}}$ & $b_{\si{J}}$ & $r_{\si{F}}$ & $M_B$ & $M_{\star}$\\
& & Dec \\
& & (J2000) & (km s$^{-1}$) & (Jy km s$^{-1}$) & (mJy) & ($10^8$ M$_{\sun}$) & (mag) & (mag) & (mag) & ($10^9$ M$_{\sun}$) \\
\hline
GalA & ESO534 & 22:36:06.69 & 1407 & 10.1 & 85.3 & 9.63 & 15.78 & 14.69 & -15.8 & 1.4 \\
 & -G001 & -26:18:51.5 \\
GalB & 2dFGRS & 04:14:37.41 & 2194 & 6.35 & 42.3 & 14.7 & 16.42 & 15.86 & -16.1 & 1.9 \\
 & S833Z022 & -22:48:25.1 \\
\hline
\end{tabular}
\tablefoot{ V$_{ \rm sys}$ is the systemic velocity; S$_{\rm int}$ is the observed integrated HI 21cm line flux; S$_{\rm peak}$ is the observed peak HI 21cm line flux; M$_{\rm HI}$ is the total hydrogen mass (all four determined by \citet{zwaan2013cold}; b$_{\rm J}$ and r$_{\rm F}$ are the extinction corrected b-band and r-band apparent magnitudes obtained from the Supercosmos Sky Survey; M$_{\rm B}$ is the absolute B-band magnitude; and M$_{\star}$ is the total stellar mass from \citet{zwaan2013cold}.}
\end{table*}

The bimodal distribution of galaxies in the colour-magnitude diagram is well established up to redshift $z$ $\sim $ 2 \citep{williams2009detection}. Red-sequence galaxies are generally characterized by a spheroidal shape, quiescence, no ongoing star formation (SF), and a low gas mass fraction. Galaxies in the blue cloud, in contrast, are disc dominated, star forming, and gas rich \citep[e.g.][]{blanton2009physical}. 
The red sequence shows a mass growth by at least a factor of 2 since $z$ $\sim$ 1 \citep{bell2004nearly}. This bimodality suggests that SF is quenched and galaxies evolve from the blue cloud onto the red sequence. In this evolutionary phase, they pass through the so-called green valley.

Post-starburst galaxies (or more specifically, E+A galaxies) are a useful tool to study the quenching
mechanism, as these galaxies are thought to be part of this transitional phase \citep[e.g.][]{Zabludoff1996environment, norton2001spatial, pracy2009kinematics}.

E+A galaxies are identified based on their optical spectra. 
The spectra show strong Balmer absorption lines, indicating a relatively young A-type stellar population, and only little or no [OII]~3727~\AA\;and H$\alpha$ emission. 
The combination of these characteristics suggests that the galaxies are post-starburst galaxies and that no significant SF occurred in the last $\sim$1~Gyr \citep{couch1987spectroscopic}. 
This indicates that E+A galaxies are snapshots of galaxies that move from the blue cloud onto the red sequence. 
However, it is still unclear whether this is the final transition. 
Half of the nearby E+A galaxies have  HI~21cm emission line detections \citep{buyle2006hi, zwaan2013cold}, suggesting that the galaxies may host further SF at a later time that is due to newly accessible gas reservoirs. 
The total number of known E+A galaxies is small at low redshift.

In this study, we present high spatial resolution follow-up observations of two HI-rich nearby low-mass E+A galaxies: ESO534-G001 (hereafter referred to as GalA) and 2dFGRS~S833Z022 (hereafter referred to as GalB) from \citet{zwaan2013cold}. These galaxies were identified as E+A galaxies based on their central optical spectra.
The stellar masses of these galaxies are log($M_{\star}$/M$_{\odot}) = 9.1$ and log($M_{\star}$/M$_{\odot}) = 9.3$, respectively \citep{zwaan2013cold}. These stellar masses are lower than or comparable to that of the Large Magellanic Cloud, which has a stellar mass of log($M_{\star}$/M$_{\odot}) = 9.3$ \citep{skibba2012spatial}. The absolute B-band luminosity of GaIA is $M_{\rm B} = -15.8$ mag and of GalB is $M_{\rm B} =-16.1$ mag. These are several magnitudes fainter than the characteristic break in the galaxy luminosity function, which in the B band is ~$M^{\star}_{B} = -20.6$ mag \citep{marzke1998galaxy}. Based on the mass and luminosity, these galaxies are therefore characterized as dwarf galaxies.

The distance to these galaxies is 20 Mpc for GalA and 31 Mpc for GalB, which is sufficiently nearby to study the gas properties with high physical resolution. The careful selection process, which restricts the observations to the southern hemisphere, is described in detail by \citet{zwaan2013cold}. The atomic hydrogen mass is determined from previous single-dish observations to be 9.63$\times 10^8$ M$_{\sun}$ for GalA and 14.7$\times 10^8$ M$_{\sun}$ for GalB \citep{zwaan2013cold}. We present high spatial resolution HI 21cm line observation and CO(1-0) emission line observations as a tracer of molecular hydrogen. Furthermore, optical integral field unit (IFU) spectra were obtained \citep{pracy2014integral}. In these IFU spectra, \citet{pracy2014integral} observed that the E+A signature is restricted to the central region and that SF is found at larger radii. Furthermore, extended UV emission is seen in the GALEX UV images. The total NUV-r colour of the galaxies is $1.5\pm 0.2$ and $1.7\pm 0.2$ for GalA and GalB, respectively. In the total NUV-r versus stellar mass plane this places these galaxies in the blue cloud. The centre, however, behaves differently from the rest of the galaxy. We did not use the GALEX data for our further analysis because of a lack of a solid estimate of dust extinction. 

The goal is to determine whether the SF activity of these galaxies is consistent with the activity seen in normal spiral galaxies, or if it shows signs of a transition towards another evolutionary phase.

This paper is structured as follows. In Sect. 2 we present newly obtained observations of neutral hydrogen using the NRAO Karl G. Jansky Very Large Array (VLA) and observations of \mbox{CO(1-0)} using the Atacama Large Millimeter and Submillimeter Array (ALMA). Additionally, a brief summary of the integral field unit spectra presented by \citet{pracy2014integral} is given. The analysis of these data sets is shown in Sect. 3. In Sect. 4 we summarize the results and discuss our conclusions. For distance-related quantities, we assume \mbox{$H_0 = 70$ $\si{km\; s^{-1}\; Mpc^{-1}}$}. We use the following conversion from angular to physical size scales: $1\arcsec \approx 97$~pc (GalA) and $1\arcsec \approx 150$~pc (GalB).

\section{Data}  

        \begin{table*}
                \caption{\label{TabObsPropVLA}Details of the VLA and ALMA observations}
                \centering
                \begin{tabular}{llcccr}
                        \hline\hline
                        & Name & Instrument & Tracer & Obs. date & Integration time\\
                        \hline
                        GalA & ESO534-G001 & VLA CnB & HI & 20.05.2012 & 13240 s\\
                                 && VLA CnB & HI & 08.05.2012 & 13240 s\\
                                 && ALMA & CO(1-0) & 13.12.2013 & 3100 s\\
                                 && VLA BnA & HI & 01.02.2014 & 20390 s\\
                                 && ALMA & CO(1-0) & 23.03.2014 & 3022 s\\
                        GalB & 2dFGRS~S833Z022 & VLA CnB & HI & 06.05.2012 & 13080 s\\
                                 && ALMA & CO(1-0) & 17.12.2013 & 2614 s\\
                                 && VLA BnA & HI & 26.01.2014 & 20200 s\\ 
                        \hline
                \end{tabular}
        \end{table*}

We present newly obtained high spatial resolution VLA and ALMA line observations of the HI 21~cm and the CO $J=1{-}0$ ($v_{\rm rest}=115.27$~GHz) line of the two selected 
HI-rich galaxies from \citet{zwaan2013cold} that have an E+A like spectrum in their central region. In our analysis we also make use of the optical IFU data that were presented by \citet{pracy2014integral}. 
These latter observations showed that the typical E+A signature is only seen in the centre of the two galaxies, whereas clumpy SF was detected at larger radii. A summary of the galaxy properties is given in Table~\ref{TabGalProp}. The details of the observations are discussed below and summarized in Table~\ref{TabObsPropVLA}. 

        \subsection{HI 21cm line observations}
        
                To study the three-dimensional structure of the atomic gas, we observed the two galaxies in the HI 21 cm emission line using the VLA in its hybrid BnA and CnB configurations. In this configuration the northern arm of the array is in a more extended setup, so as to observe targets at low declination with a nearly circular beam.  The observations were carried out using the WIDAR correlator in dual polarization mode, giving 7.8~kHz wide channels. The resulting channel separation is $\sim$1.7~km~s${^-1}$ and the resolution after Hanning smoothing is $~$3.3~km~s${^-1}$. The total bandwidth of 4~MHz translates into a total velocity range of $~$860~km~s$^{-1}$, sufficiently wide to include all the HI emission, as can be seen from the Green Bank Telescope (GBT) spectra.  The details of the observations are summarized in Table~\ref{TabObsPropVLA}. 

The calibration and data reduction was carried out using the Common Astronomy Software Applications (CASA 4.2.2) software package.  The editing was done interactively, allowing us to remove parts of the data that suffered from radio frequency interference (RFI) or other obvious technical issues. The calibration was done using appropriate flux and bandpass calibrator sources and standard procedures. The time-variable complex gain was calibrated using frequent observations of a standard unresolved calibrator source in the same region on the sky. The flux scale was set using an observation of the source 3C147. Before imaging, the continuum emission was removed from the $u,v$ data set by fitting a constant to the real and imaginary parts of the line-free channels.

The imaging was done using the standard clean algorithm. The multi-scale option was used, which is optimized for high spatial resolution observations of extended sources. A 'robust' weighting scheme was applied, using a Briggs parameter of 0.5, which guarantees a nearly optimal sensitivity while still providing a high spatial resolution and a well-behaved synthesized beam. The resulting mean r.m.s. noise level is measured to be $\sim$0.7~mJy~beam$^{-1}$ per 3.3~km~s$^{-1}$. The angular resolution of the HI data cube is $\sim$4\arcsec, and the largest angular scale observable is $\sim$16\arcmin.

To ensure that only real emission is taken into account in the subsequent analysis, we define masking regions based on lower resolution data cubes. To that end, we use only the data from observations in the CnB configuration and convolve the resulting image cube with a Gaussian kernel with a full width at half maximum of 25\arcsec. We define the masking regions as those regions that correspond to HI emission above a 2~$\sigma$ noise level. For the moment analysis of the high-resolution data cubes, these masking cubes are used to separate real emission from noise.

        \subsection{CO(1-0) observations}
We use the CO(1-0) emission line at rest frequency of 115.27~GHz to trace the molecular hydrogen content and kinematics of our sample galaxies. ALMA Cycle~1 observations in Band~3 were obtained, the characteristics of which are summarized in Table~\ref{TabObsPropVLA}. A phase calibrator was observed every $\sim$8 minutes, and a bandpass calibrator was observed for 15 minutes at the beginning of the observations. The absolute flux scale was obtained through observing unresolved targets whose flux scale is regularly monitored by the ALMA observatory.  The total on-source integration times were 100 minutes for GalA and 40 minutes for GalB. The data were taken with four basebands, one with a velocity resolution of 2.5~km s$^{-1}$ centred on the expected CO line, and the other three in low-resolution mode at lower frequencies. For these observations, 32 antennas were available, but 5 of these  had to be flagged during the data reduction procedure. The full reduction
of the VLA data was carried out with CASA. 

Cubes were made using a slight tapering of the longest baseline data and a 'robust' weighting scheme with a Briggs parameter of 0.5 to obtain a spatial resolution of 3~arcsec. The velocity resolution of the cubes was set at 10~km s$^{-1}$.  The largest observable scale with the array configuration we used is $\sim$20\arcsec. The measured r.m.s. noise level is $\sim$1.5~mJy~beam$^{-1}$ per 10~km s$^{-1}$.

        \subsection{Integral field spectra}
        
                \citet{pracy2014integral} present the optical IFU observations used in this study. 
                The target galaxies were observed on 2012 November 12 using the  Wide Field Spectrograph (WiFeS).
                It is operated on Mount Stromlo with the 2.3m
telescope
of the Siding Spring Observatory. The average seeing was 1.7\arcsec.
                The spaxel size is 1\arcsec~x~1\arcsec~and the total field of view is 25\arcsec~x~38\arcsec.
                The spectra span a wavelength range from $\sim$3500~to~9000~\AA~with a spectral resolution of R$\sim$3000. 
                 In this paper, we use the fully reduced and flux-calibrated data cubes from \citet{pracy2014integral}.

\section{Analysis and results}

                \begin{figure*}
                \includegraphics[width=0.455\linewidth]{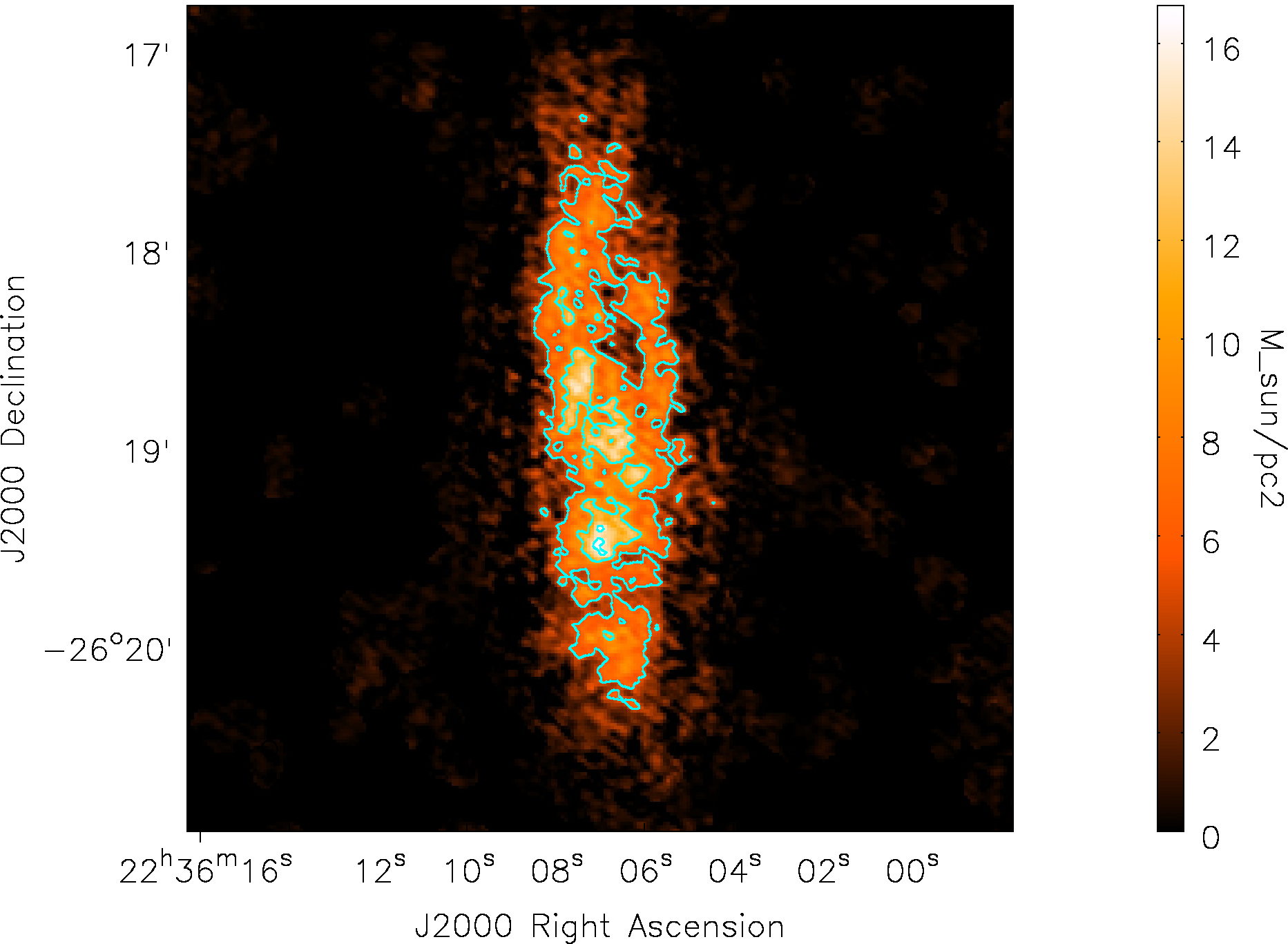}\textbf{(a)} \hfill
                \includegraphics[width = 0.455 \linewidth]{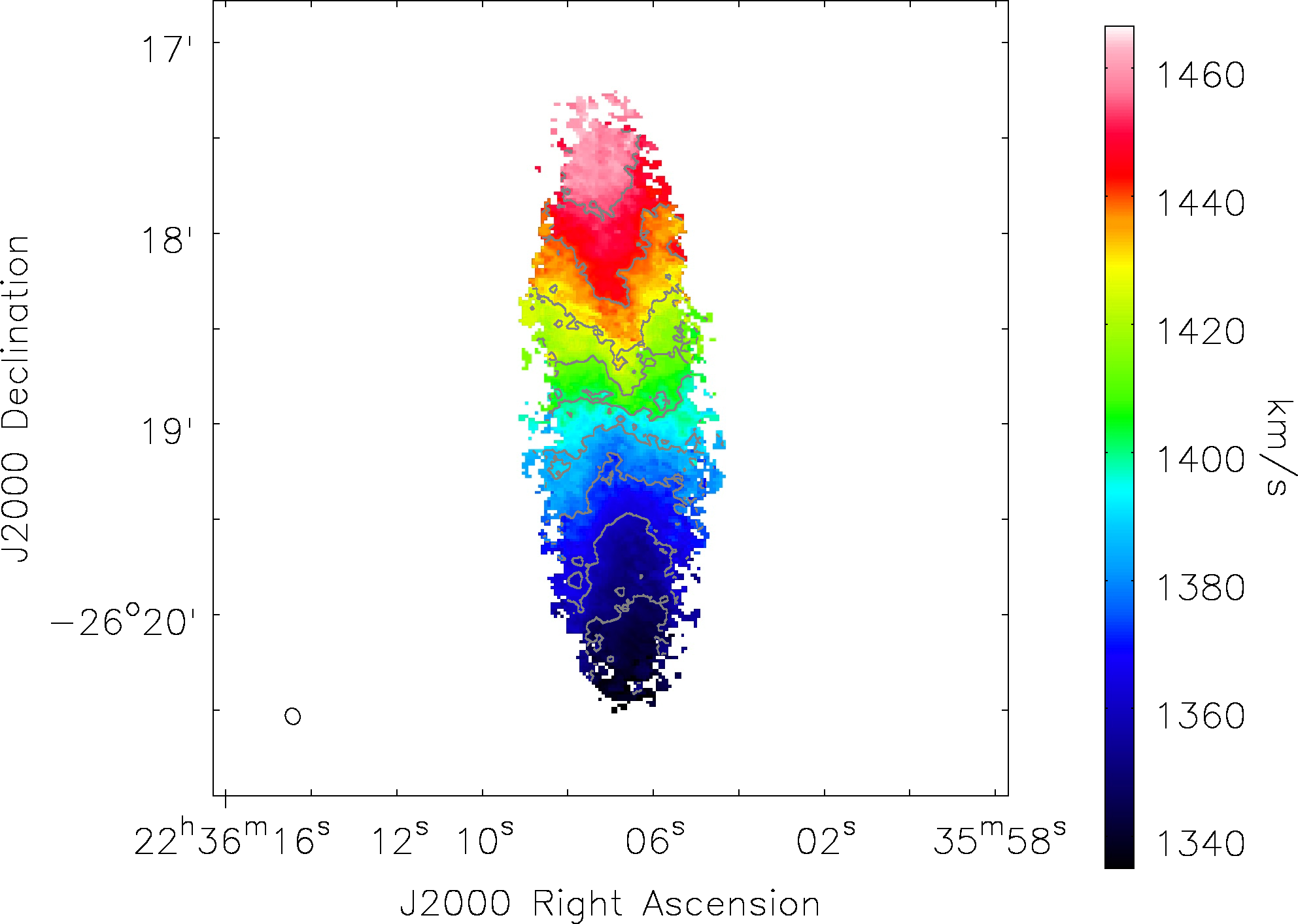}\textbf{(b)} \\
                \includegraphics[width = 0.455 \linewidth]{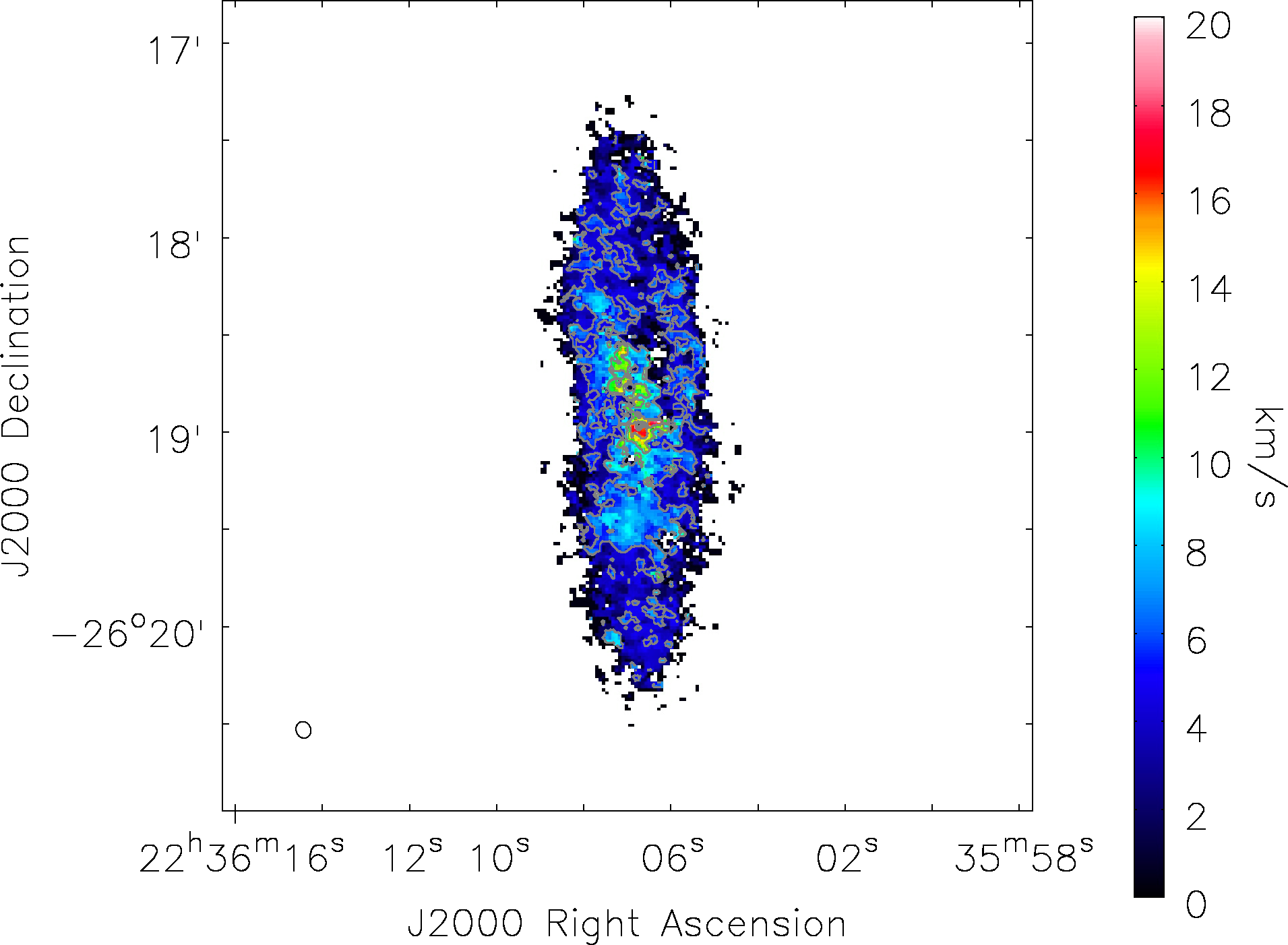}\textbf{(c)}  \hspace{0.8cm}
                \includegraphics[width=0.35\linewidth]{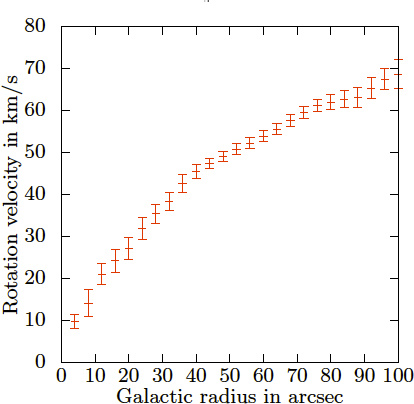}\textbf{(d)} \\
                \includegraphics[width=0.455\linewidth]{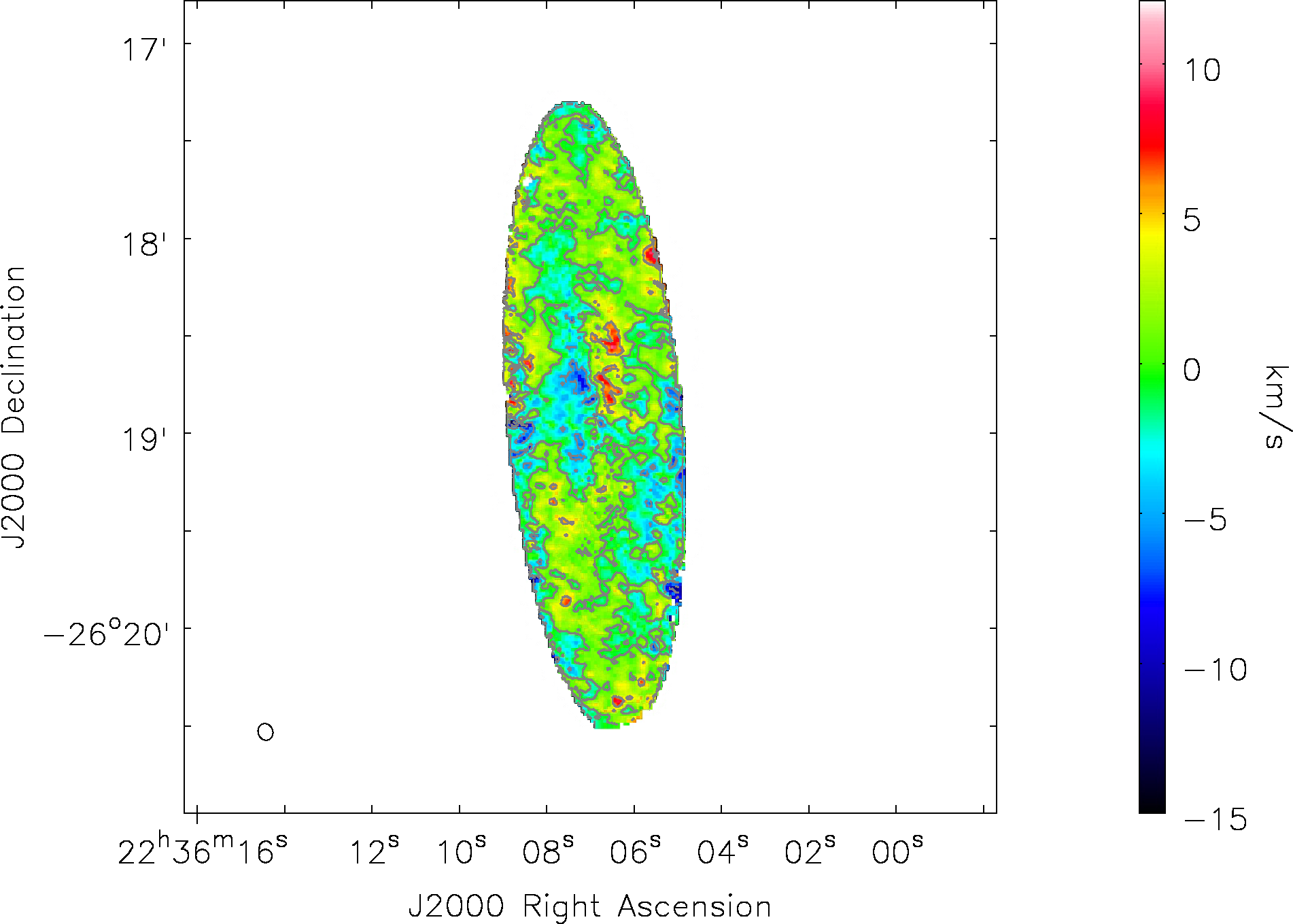}\textbf{(e)} \hfill
                \includegraphics[width=0.455\linewidth]{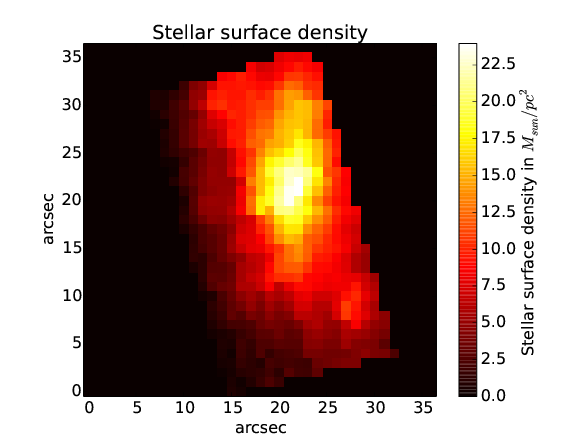}\textbf{(f)}  
                \caption{Results for GalA. \textbf{(a):} HI surface density in $M_{\sun}$ pc$^{-2}$ from the combined VLA observations. The spacing between the contours is 5~M$_{\sun}$ pc$^{-2}$. \textbf{(b):} Line-of-sight velocity map of HI observed using the VLA. The $\Delta v$ spacing between the contours is 14~km~s$^{-1}$. \textbf{(c):} Velocity dispersion map. The spacing between the contours is 5~km~s$^{-1}$. \textbf{(d):} Rotation curve of the HI disc observed with the VLA. The curves are obtained from the tilted ring model described in the text. \textbf{(e):} HI residual velocity field. The contours are at -10, 0 and 10 km s$^{-1}$. \textbf{(f):} Stellar surface density map obtained from the B-band luminosity determined from IFU observations and a colour-dependent mass-to-light ratio \citep{bell2003optical}. In these maps, north is at the top and east is on the left-hand side.}
                \label{FigAnalysisResultsESO}
        \end{figure*}
        \begin{figure*}
                \includegraphics[width=0.455\linewidth]{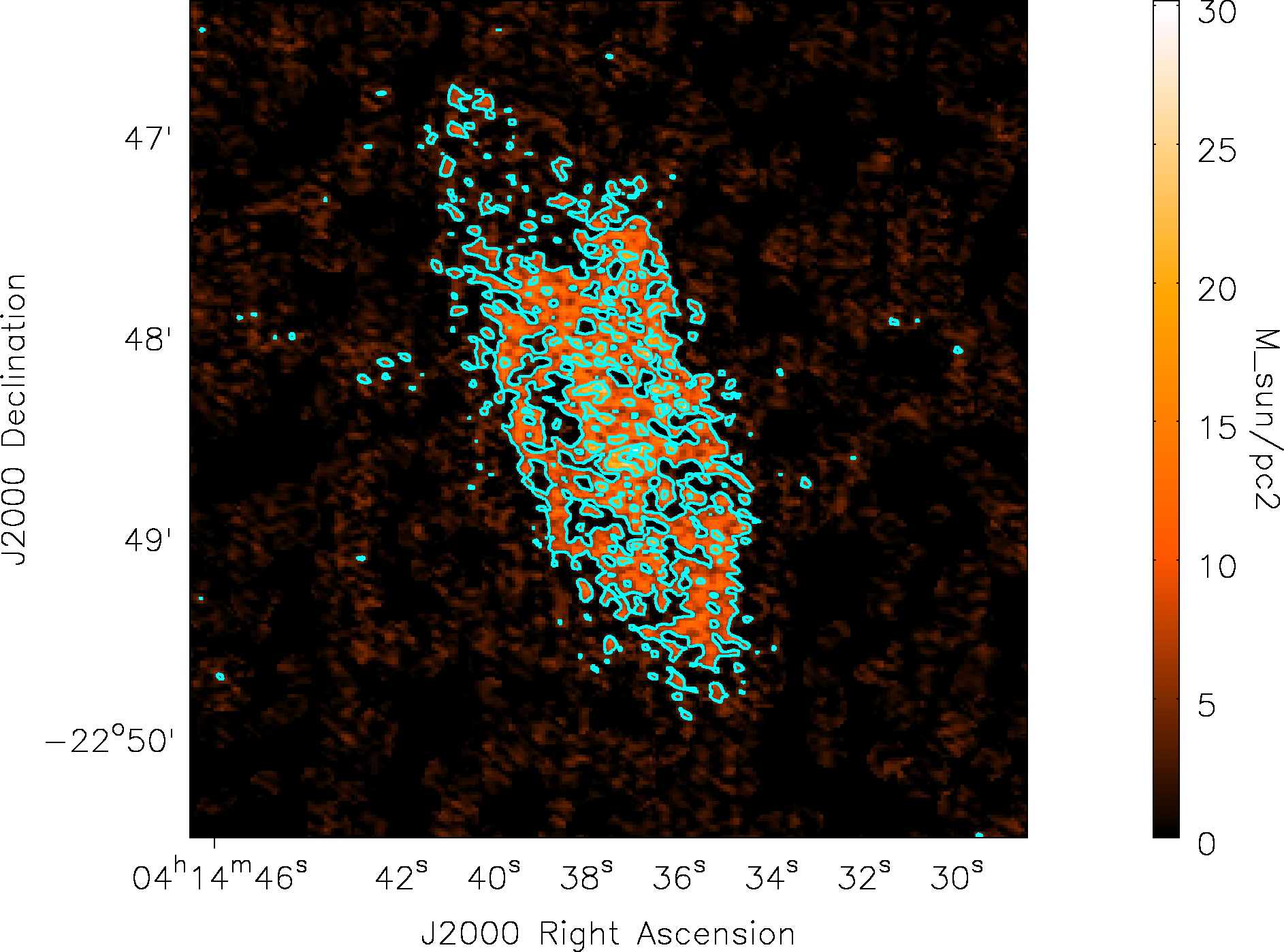}\textbf{(a)}  \hfill
                \includegraphics[width = 0.455 \linewidth]{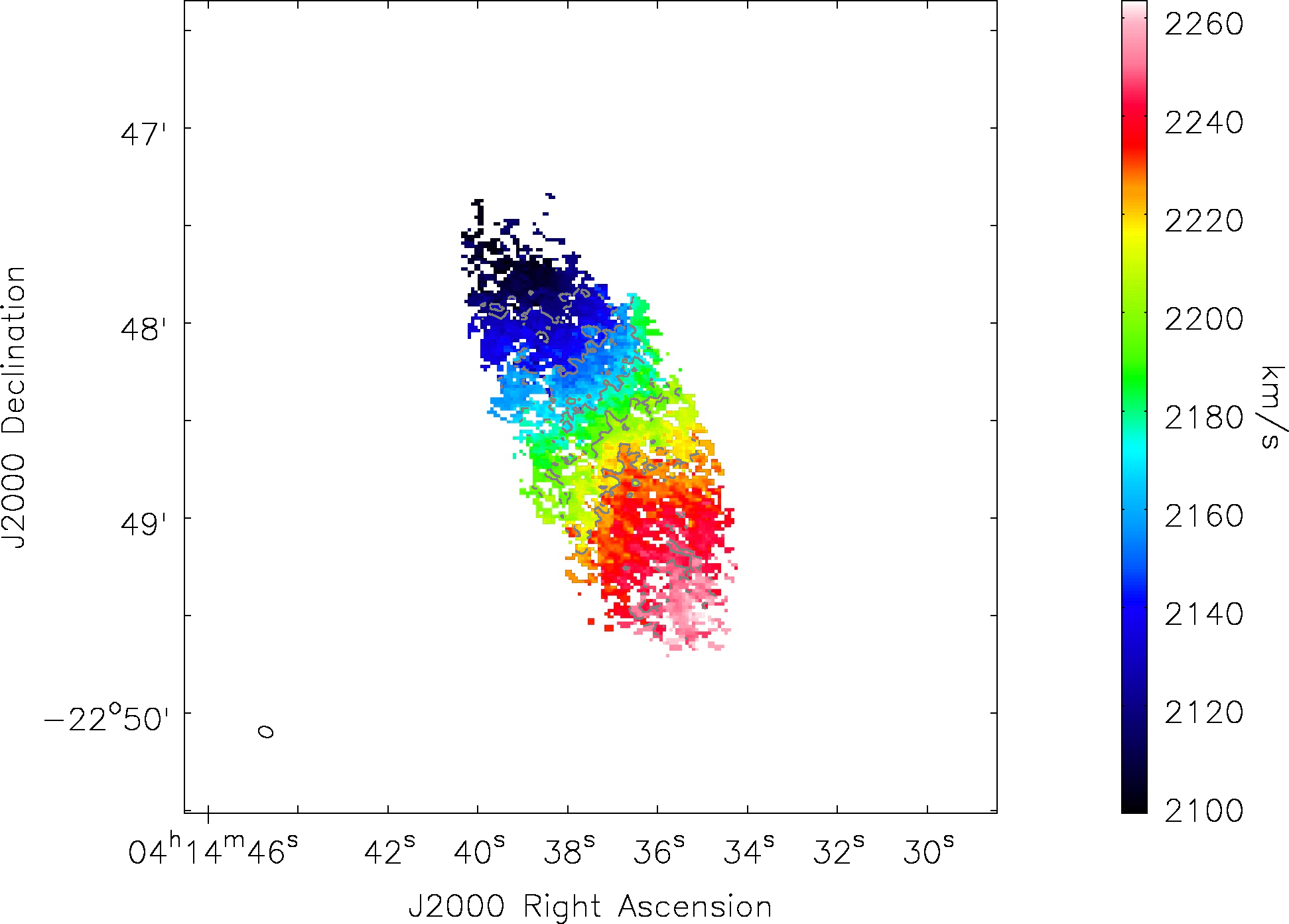}\textbf{(b)} \\
                \includegraphics[width = 0.455 \linewidth]{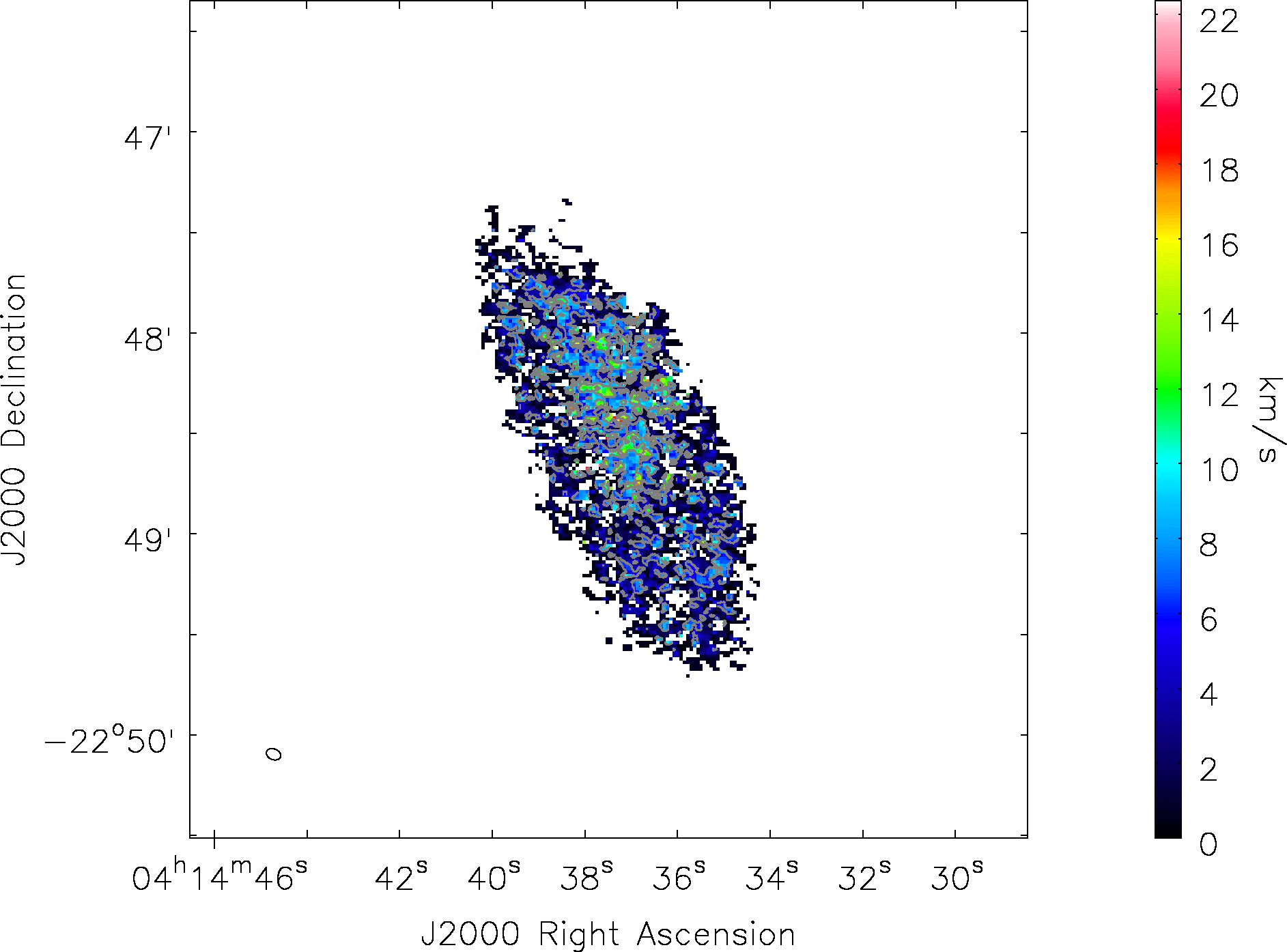}\textbf{(c)}  \hspace{0.7cm}
                \includegraphics[width=0.35\linewidth]{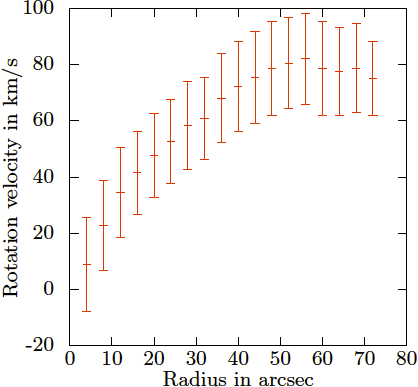}\textbf{(d)} \\
                \includegraphics[width=0.455\linewidth]{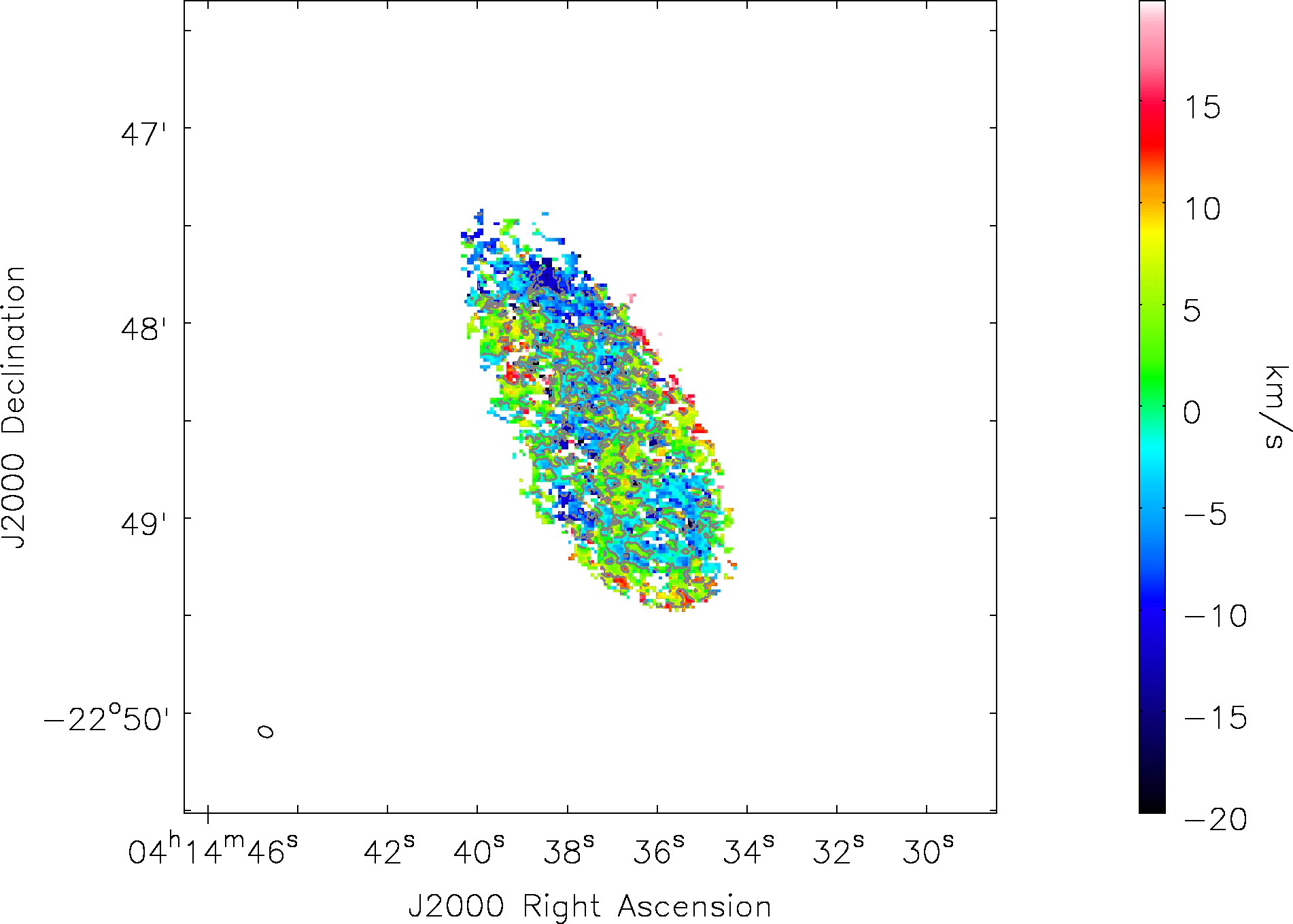}\textbf{(e)}  \hfill
                \includegraphics[width=0.455\linewidth]{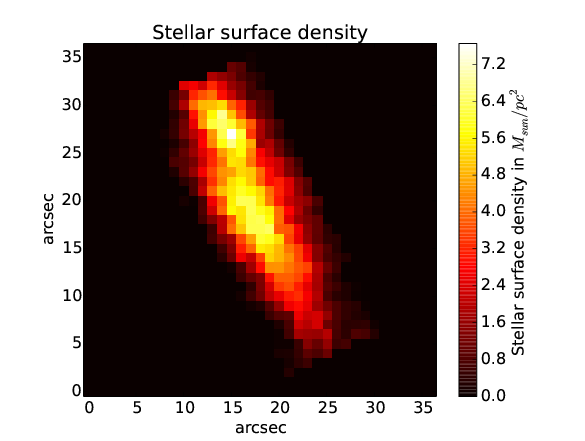}\textbf{(f)} 
               \caption{Analysis results for GalB. For a description of each plot see the caption of Fig.~\ref{FigAnalysisResultsESO}. The differences with Fig.~\ref{FigAnalysisResultsESO} are that for the HI surface densities the contours are at 5, 15, and 25 ~M$_{\sun}$ pc$^{-2}$. For the line-of-sight velocity the $\Delta v$ spacing between the contours is  26~km~s$^{-1}$.}
                \label{FigAnalysisResults2dF}
        \end{figure*}

        \begin{table*}
                \caption{\label{TabGalPropGas}Summary of galaxy gas properties determined from the HI 21cm line observations and the CO(1-0) emission line observations presented in this work.
                }
                \centering
                \begin{tabular}{llccccccc}
                        \hline\hline
                        &Name  & i & $V_{\si{sys}}^{\si{HI}}$ & $S_{\si{int}}^{\si{HI}}$ & S$_{\si{peak}}^{\si{HI}}$ & M$_{\si{HI}}$ & $L_{\rm CO}$\\
                        &        & ($^{\circ}$) & (km s$^{-1}$) & (Jy km s$^{-1}$) & (mJy)  & (10$^8$M$_{\sun}$) & (10$^4$ K km s$^{-1}$ pc$^2$)\\
                        \hline
                        GalA & ESO534-G001 & 74 & 1400.0 & 10.68 & 101.7 & 10.1 & $<$1.9\\
                        GalB & 2dFGRS~S833Z022 & 60 & 2194.0 & 5.01 & 56.86 &  11.27 & $<$4.6\\
                        \hline
                \end{tabular}
                \tablefoot{ $i$ is the inclination angle measured from the HI 21cm line observations, $V_{\si{sys}}^{\si{HI}}$ is the systemic velocity determined from the tilted ring model, $S_{\si{int}}$ is the integrated  intensity, S$_{\si{peak}}$ is the peak intensity, M$_{\si{HI}}$ is the total HI mass, and $L_{\rm CO}$ is the integrated CO luminosity.
                }
        \end{table*}

        \subsection{Neutral hydrogen}

                From the 21~cm line observations we infer the HI distribution as well as the kinematics.      
                The HI mass surface density is calculated by integration of the spectrum over the frequency axis. 
                High spatial resolution surface density maps are shown in Figs.~\ref{FigAnalysisResultsESO}a and \ref{FigAnalysisResults2dF}a. 
                For the conversion from spectral flux densities to HI mass surface densities, we use
                \begin{equation}
                        \Sigma_{\si{HI}}\;[\si{M}_{\odot} \si{pc^{-2}}]= 0.020 \cos i \; I\;[\si{K \; km \; s^{-1}}]
                \end{equation}
                deprojecting and including a factor of 1.36 to account for the presence of helium \citep{leroy2008star}.
                The gas mass surface densities have peak values of 16.79 M$_{\sun}$ pc$^{-2}$ (GalA) and 30.23~M$_{\sun}$ pc$^{-2}$ (GalB). 
                The gas is distributed relatively regularly over each galaxy with a mean HI mass surface density of 3.37~M$_{\sun}$ pc$^{-2}$ (GalA) and 5.3~M$_{\sun}$ pc$^{-2}$ (GalB), measured over the entire observed HI discs.
                The total neutral hydrogen masses we determine are $M_{\si{HI}}$ = (10.1~$\pm$~0.2)~$\times10^8$M$_{\sun}$ (GalA) and $M_{\si{HI}}$ = (11.3~$\pm$~0.6)~$\times10^8$~M$_{\sun}$ (GalB).
                A summary of the properties derived from the surface density maps is given in Table~\ref{TabGalPropGas}. 
                These results are consistent within the uncertainties with the results from the single-dish observations that are listed in Table~\ref{TabGalProp} based on \citet{zwaan2013cold}, where lower observed fluxes result from the fact that only flux on certain scales is visible with an interferometer. 

                The rotational velocity is determined based on the intensity-weighted mean velocity maps, which are shown in Figs.~\ref{FigAnalysisResultsESO}b and \ref{FigAnalysisResults2dF}b. 
                For the creation of these maps an additional 3$\sigma$ noise level cut is applied to the masked data cube to ensure a high signal-to-noise ratio (S/N). 
                The observed line-of-sight velocities are consistent with a rotating disc.
                To determine the deprojected rotation curves, we use the tilted ring model of \citet{begeman1989hi}, which describes a galaxy as a set of concentric rings with the following parameters: position of the centre ($x_0$, $y_0$), systemic velocity $V_{\si{sys}}$, rotation velocity $v_{\si{rot}}$, inclination $i$, position angle (PA) measured anti-clockwise between north on the sky and the semi-major axis of the receding part of the galaxy.
                For the application of the tilted ring model, we use the \textsc{GIPSY} task \textsc{ROTCUR}. 
                We subsequently fix the parameters in a four-step procedure following \citet{noordermeer2007mass}.
                In this way, we can infer the radial dependence of the parameters and exclude warps in the discs, which would be observed as radius-dependent inclination or position angle. 
                The resulting rotation curves are plotted in Figs.~\ref{FigAnalysisResultsESO}d and \ref{FigAnalysisResults2dF}d. 
                An approximately constant slope is observed in the inner part for both galaxies, which is consistent with a rotating solid body. 
                At larger radii the slope of the rotation curves starts to decrease. 
                For GalB in Fig.~\ref{FigAnalysisResults2dF}d we even observe a flat rotation curve at large radii. 
                For GalA in Fig.~\ref{FigAnalysisResultsESO}d, however, a flat rotation curve is not reached within the maximum radius that can be observed. 
                For dwarf galaxies, the flat part of the rotation curve is not necessarily seen, as we do not observe the HI disc to large enough radii.   \citep{swaters2009rotation}.
                The rotation velocity increases to (69~$\pm$~3)~km~s$^{-1}$~(GalA) and (75~$\pm$~13)~km~s$^{-1}$~(GalB), where we use the dispersion of velocities along each tilted ring as the error of the rotation velocity. 
                
                The accuracy of the model velocity field is assessed from the residual velocity field, which is defined as the difference between observed and model velocity field. 
                The residual velocity maps are shown in Figs.~\ref{FigAnalysisResultsESO}e~and~\ref{FigAnalysisResults2dF}e. 
                The plots demonstrate that there are no systematic deviations. 
                The mean deviation is (0~$\pm$~3)~km~s$^{-1}$ for GalA and (1~$\pm$~7)~km~s$^{-1}$ for GalB.

                The velocity dispersion is determined based on the intensity-weighted velocity dispersion maps
                , which are shown in Figs.~\ref{FigAnalysisResultsESO}c and \ref{FigAnalysisResults2dF}c. 
                For the calculation of these maps, the same 3$\sigma$ cut is applied as for the rotational velocity maps. 
                The velocity dispersion increases to maximum values of 20~km~s$^{-1}$ (GalA) and 22~km~s$^{-1}$ (GalB) in the galaxy centres. 
                Overall, the velocity dispersion decreases from the centre to larger radii, which is most clearly visible for GalA. 
                The mean velocity dispersion observed over the entire observed HI discs is 4.4~km~s$^{-1}$ (GalA) and 4.6~km~s$^{-1}$ (GalB). 
                These values are on the low end of the range observed in nearby star-forming spiral galaxies \citep{walter2008things, ianjamasimanana2012shapes} and dwarf galaxies \citep{hunter2012little}.

        \subsection{Molecular hydrogen}
        
        \begin{figure}
        \begin{center}
        \includegraphics[trim = 2 3 3 3, width=\linewidth]{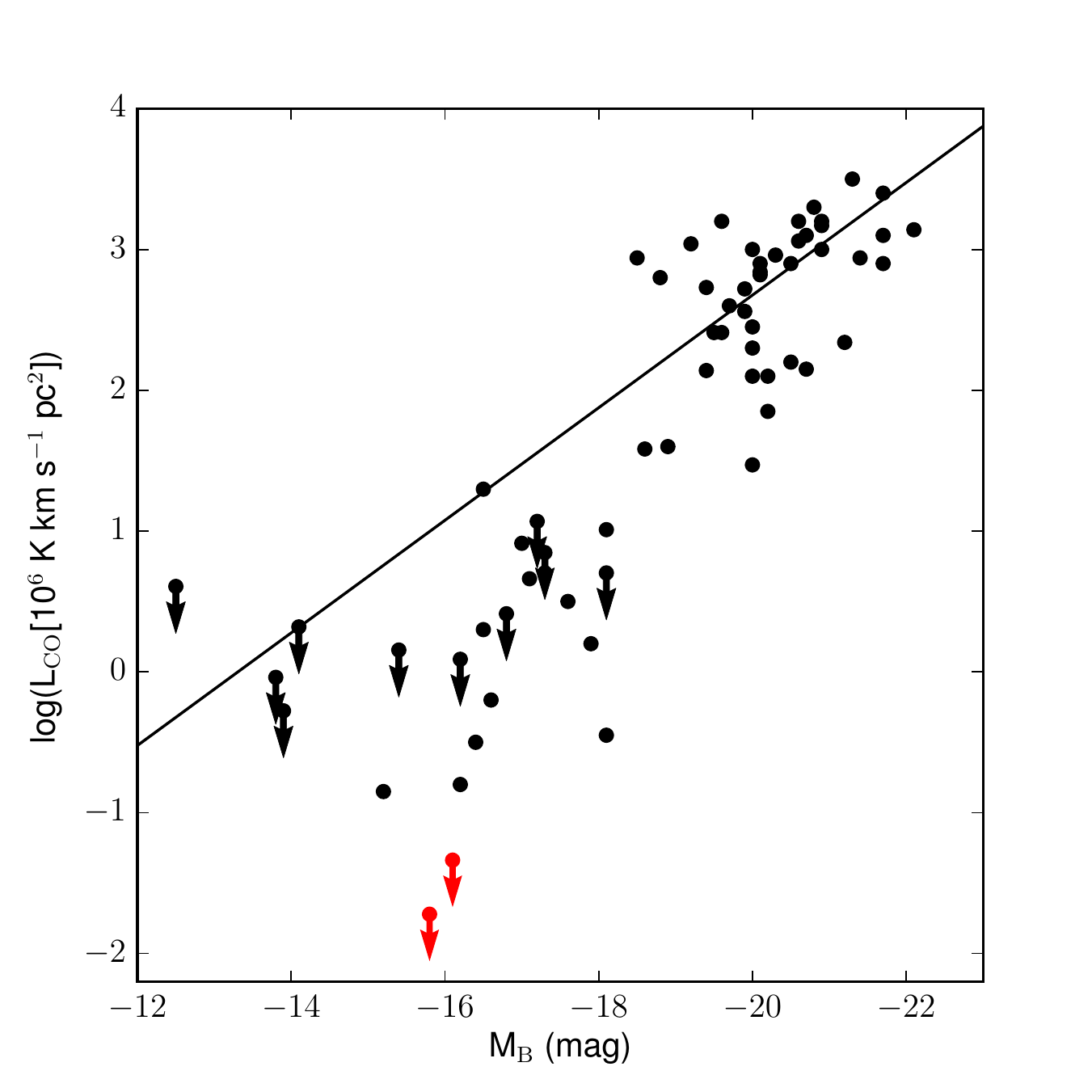}
        \end{center}
        \caption{Total CO(1-0) luminosity as a function of B-band luminosity for normal spirals and dwarf galaxies presented by \citet{schruba2012low} in black and our galaxies in red. The black line is a linear fit to the bright galaxies (M$_{\rm B}$ $<$ -18) also taken from \citet{schruba2012low}. }
        \label{FigCompSchruba}
        \end{figure}

                After scrutinising the ALMA CO(1-0) cubes, we do not detect any CO(1-0) emission in either of the two galaxies. Therefore, we can only determine upper limits to the total CO(1-0) flux, and hence to the total molecular hydrogen present in the two galaxies. We note here that the ALMA observations that we obtained are only sensitive to observable scales below $\sim$ 20\arcsec, which correspond to physical scales of {1.94}~kpc for GalA and {3.01}~kpc for GalB. Any coherent emission on spatial scales larger than these values would generallly escape detection by our observations.

        For the calculation of upper limits, we stack CO spectra over as large a region as possible in an attempt to enhance the S/N. To this end, we first de-rotate the CO(1-0) data cubes, making sure that all CO spectra are aligned in velocity space, with zero-velocity being equal to the local radial component of the rotational velocity. We use the measured HI  velocity at each pixel and shift all the corresponding spectra accordingly. To set the spatial region over which we average, we use as a criterion an HI surface density exceeding a threshold of 1~M$_{\sun}$~pc$^{-2}$. The CO spectra at all  positions fulfilling that criterion are then added up to a total spectrum from which we determine the upper limit to the integral CO flux. To calculate this total flux, we sum the spectra over a velocity range corresponding to plus or minus the measured HI  velocity dispersion. Since CO velocity dispersions in galaxy discs are typically lower than those measured in HI \cite[see][]{mogotsi2016hi}, this choice of velocity limits ensures that no CO flux is missed. The uncertainty on the flux measurement is calculated by repeating the same exercise 60 times, but each time offsetting the central velocity by at most 300 km s$^{-1}$.

                Using this procedure, we evaluate 3$\sigma$ upper limits to the total CO(1-0) luminosity of $L_{\si{CO}}=1.9 \; \si{K \; km \; s^{-1} \; pc^2}$ and $4.6 \; \si{K \; km \; s^{-1} \; pc^2}$ for GalA and GalB, respectively. We convert these values into total molecular gas masses using the following equation:

        \begin{equation}
                M(\si{H}_2) = \alpha_{\si{CO}} \times L_{\si{CO}},
                \end{equation}
        where $\alpha_{\rm CO}$ is the conversion factor between the CO line intensity and the molecular gas surface density (see below).

                The CO-to-H$_2$ conversion factor is strongly dependent on the metallicity. For low metallicities, \citet{schruba2012low} found that $\alpha_{\rm CO}$ increases by more than an order of magnitude.
                
                Using the spatially integrated spectra from \cite{pracy2014integral}, we determine the gas metallicity with the R23 method \cite[see e.g.][]{pagel1979composition, kewley2008metallicity}:
                \begin{equation}
                \text{R}_{23} = ([\text{OII}] + [\text{OIII}])/\text{H}_{\beta}.
                \end{equation}
                We correct the spectra for stellar contribution by fitting and removing an optimal template from the MILES library \citep[v9.1;][]{sanchez-blazquez2006medium} with \texttt{ppxf} \citep{cappellari2004parametric}. Then, following \cite{dominguez2013dust}, we calculate $E(B-V)=1.97 \log_{10}[(\text{H}_{\alpha}/\text{H}_{\beta})_{obs}/2.86]$ assuming Case B recombination \citep{osterbrock1989astrophysics} from the measured Balmer decrement. The galaxy spectra are then de-reddened with the \cite{calzetti2000dust} procedure.

                We measure $\log(R23) = 0.65$ and $0.73$ for GalA and GalB, respectively. To determine whether we are on the lower or upper branch of R23, we also determine the $\log([\text{NII}]/[\text{OII}])$ and $\log([\text{NII}]/[\text{H}_{\alpha}])$ values. They amount to $-1.27$ and $-0.91$ for GalA and $-1.24$ and $-1.18$ for GalB. The [NII]/[OII] ratios indicate the upper branch of R23, whereas for [NII]/H$_\alpha$ we are in the intermediate regime. We assume the upper branch in the metallicity estimates.

                Using the \cite{kobulnicky2004metallicities} equations as described in \citet[Appendix A2.3,]{kewley2008metallicity} we calculate a gas metallicity of $12 + \log(\text{O}/\text{H}) = 8.81$  and $8.81$ for GalA and GalB, respectively. However, if we use the equations by \cite{pettini2004OIII/NII} as described in \citet[Appendix A4]{kewley2008metallicity} for the N2 ratio, we obtain $12 + \log(\text{O}/\text{H}) = 8.13$  and $8.16$ for GalA and GalB, respectively. Given the large uncertainty in measuring the gas metallicity, we adopt $12 + \log(\text{O}/\text{H}) = 8.2$ for this paper.

                Based on the work by \citet{elmegreen2013carbon}, we estimate that $\alpha_{\rm CO} \approx 70 \; \si{M}_{\odot}\si{(K \; km \; s^{-1} \; pc^2)^{-1}}$ is an appropriate conversion factor. This would lead to an upper limit for the molecular hydrogen mass of 1.3~$\times10^6$~M$_{\sun}$ (GalA) and $3.3 \times 10^6$~M$_{\sun}$ (GalB). These masses are extremely low compared to molecular hydrogen masses of $10^{7.7}$--$10^{9.8}$~M$_{\odot}$ found for E+A galaxies by \citet{french2015discovery}. However, the galaxies observed by \citet{french2015discovery} have stellar masses of log(M$_{\star}/\si{M}_{\odot}) \geq 9.96$. Thus, previous studies did detect CO(1-0) line emission in E+A galaxies that are about an order of magnitude more massive than GalA and GalB.

                Even with a value of $\alpha_{\rm CO}$ appropriate for the low metallicity of these galaxies, the molecular hydrogen mass is very low. This could be explained by the low gas densities and possibly a strong radiation field that would destroy the CO molecules. In that case, the molecular hydrogen would not be well traced by CO.
                
                Finally, we compare the upper limits in L$_{\rm CO}$ with a literature sample of normal spirals and dwarf galaxies from \citet{schruba2012low}, shown in Fig. \ref{FigCompSchruba}. In their study, \citet{schruba2012low} found that in spirals with M$_{\rm B} \leq -18$, M$_{\rm B}$ and L$_{\rm CO}$ trace each other nicely.  Dwarf galaxies are found to be under-luminous in CO compared to this relation. The two galaxies presented in this study add further evidence to this. However, most of the low optical luminosity galaxies have only upper limits in L$_{\rm CO}$. With our ALMA observations we reach a higher sensitivity and therefore a lower upper limit than \citet{schruba2012low}, but it clearly indicates that to constrain this relation for dwarf galaxies, observations of L$_{\rm CO}$ with higher sensitivity
in dwarf galaxies are needed.

        \subsection{Integral field spectra}
        
                In the original optical study the IFU was aligned with the major axis of the galaxies \citep{pracy2014integral}. 
                For this work, however, we use the rotated data cubes to have all data sets aligned. 
                The coverage of GalA is limited, but GalB is covered in its entirety.

                From the IFU spectra we derive the stellar mass surface density using a colour-dependent mass-to-light ratio \citep{bell2003optical}:
                \begin{equation}
                        \log_{10}\left(\frac{M}{L_{\lambda}} \right) = a_{\lambda} + (b_{\lambda} \times \text{colour} ),
                \end{equation}  
                where $ L_{\lambda}$ is the luminosity in a specific passband, $a_{\lambda}$ and $b_{\lambda}$ are passband-specific parameters, and the colour can be chosen among six different colours listed by \citet{bell2003optical}. 
                To determine the stellar mass surface density, we use the luminosity in the $B$ band and the $B-R$ colour. To determine the $B-R$ colour, we use the conversion given by \citet{couch1980photometric}.
                The corresponding parameters are given as $a_{B}=-1.224$, $b_{B}=1.251$.
                Figures~\ref{FigAnalysisResultsESO}f and \ref{FigAnalysisResults2dF}f show the stellar mass surface density maps for GalA and GalB, respectively.
                
                Using the spatially integrated spectra from  \citet{pracy2014integral}, we determine the stellar velocity dispersions with ppxf  \mbox{\citep{cappellari2004parametric}} and the MILES library (v9.1;  \citeauthor{sanchez-blazquez2006medium}  \citeyear{sanchez-blazquez2006medium}). The galaxies’ spectral resolution was adjusted to the MILES library \citep{beifiori2011spectral}, and emission lines were masked out in the fit. These fits result in \mbox{$\sigma = (146\pm 11)\,\si{km\, s^{-1} \, and \;} \sigma<100\,\si{km\, s^{-1}}$} for GalA and GalB, respectively. Because the spectra are dominated by the Balmer absorption lines, the fits are somewhat unstable and the error might be even larger than the above-quoted formal error. 
                
                Furthermore, we use the H$\alpha$ emission observed in the IFU spectra as a comparison for the star formation rate (SFR) predictions calculated in Sect. \ref{SubsecSFRprediction}.

\section{Star formation rate prediction} \label{SubsecSFRprediction}
                
        \begin{figure*}
                \center
                \includegraphics[width=0.4 \linewidth]{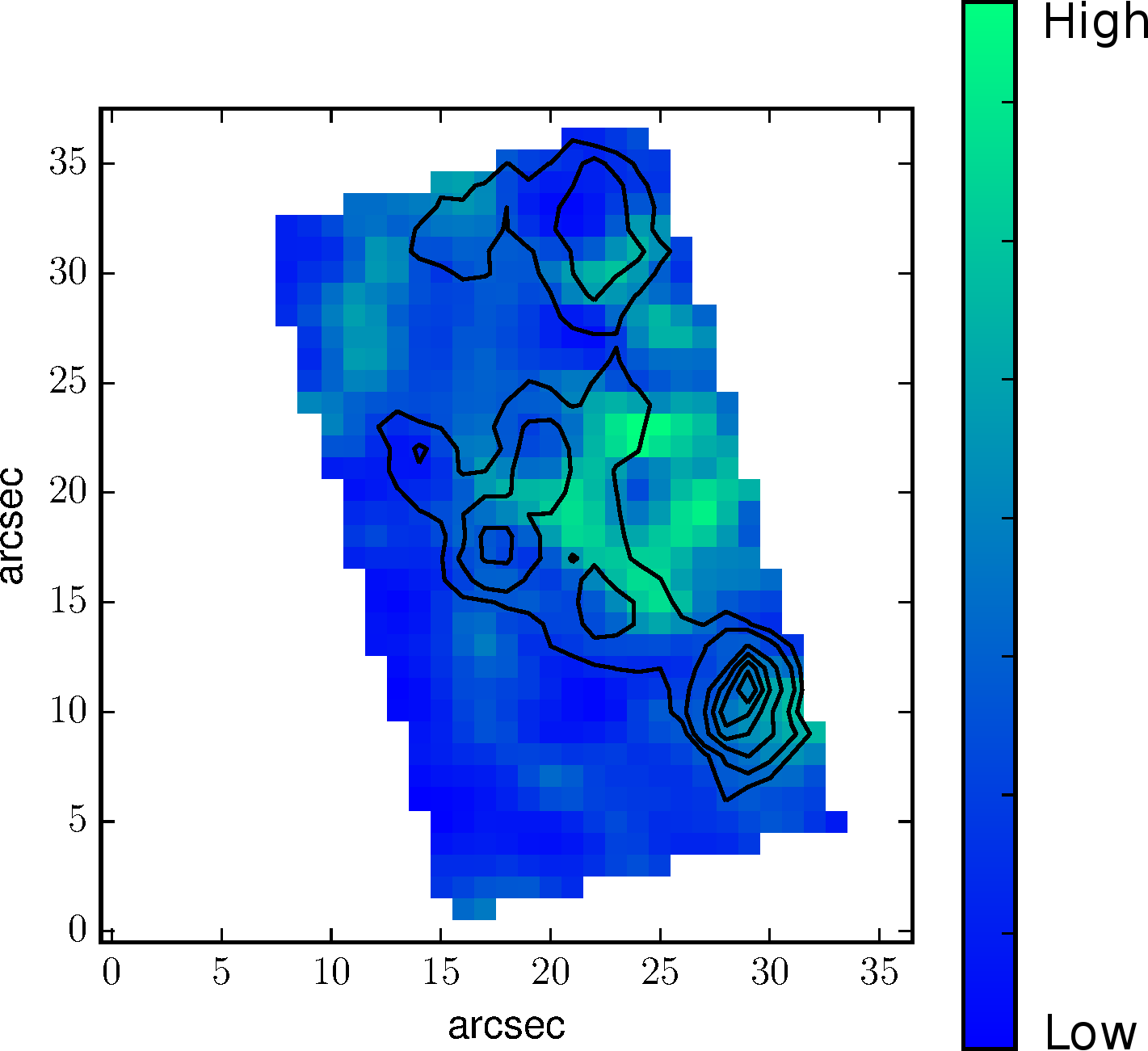}
                \includegraphics[width=0.4\linewidth]{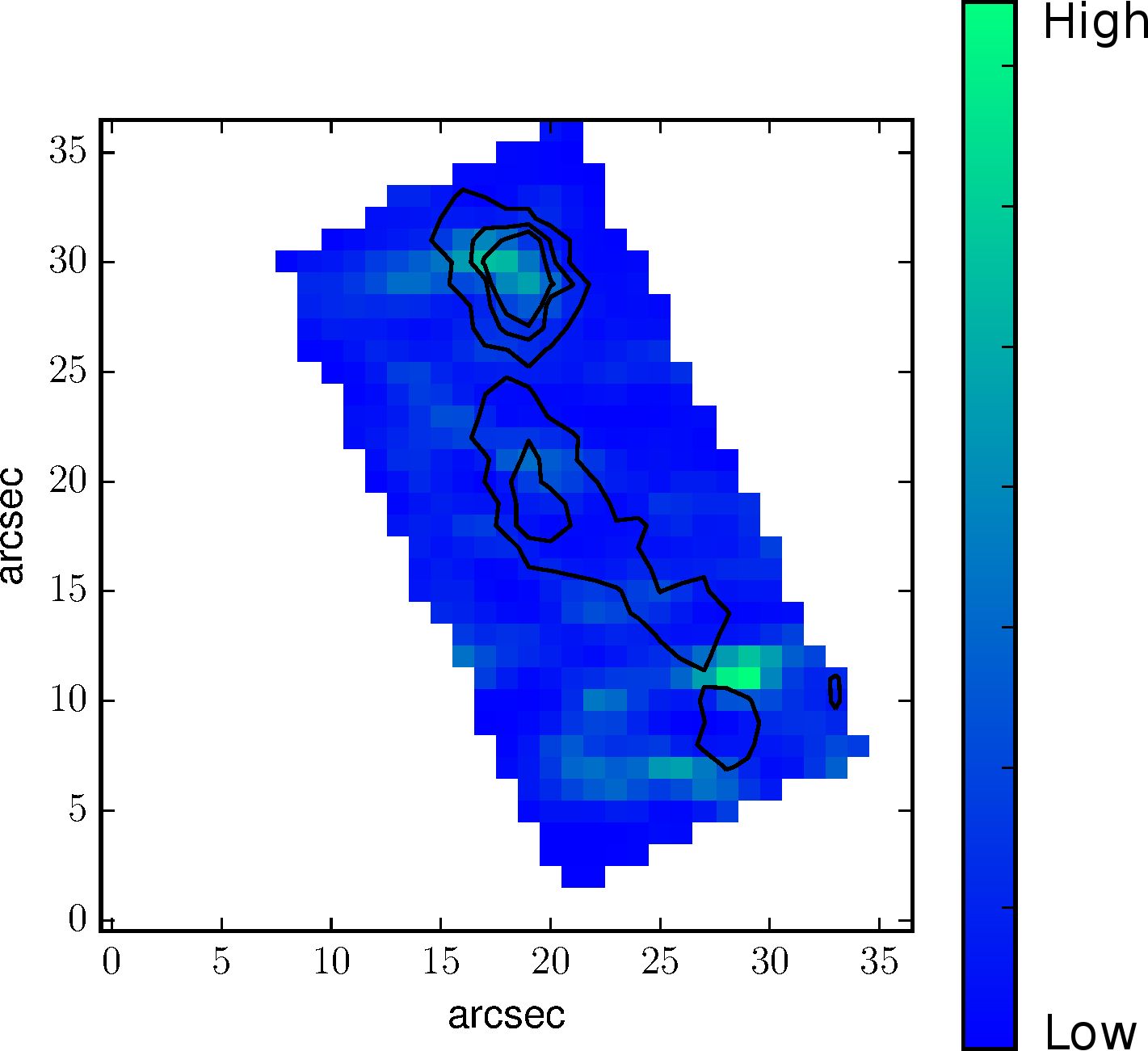} \vspace{0.5cm}

                \includegraphics[width = 0.4 \linewidth]{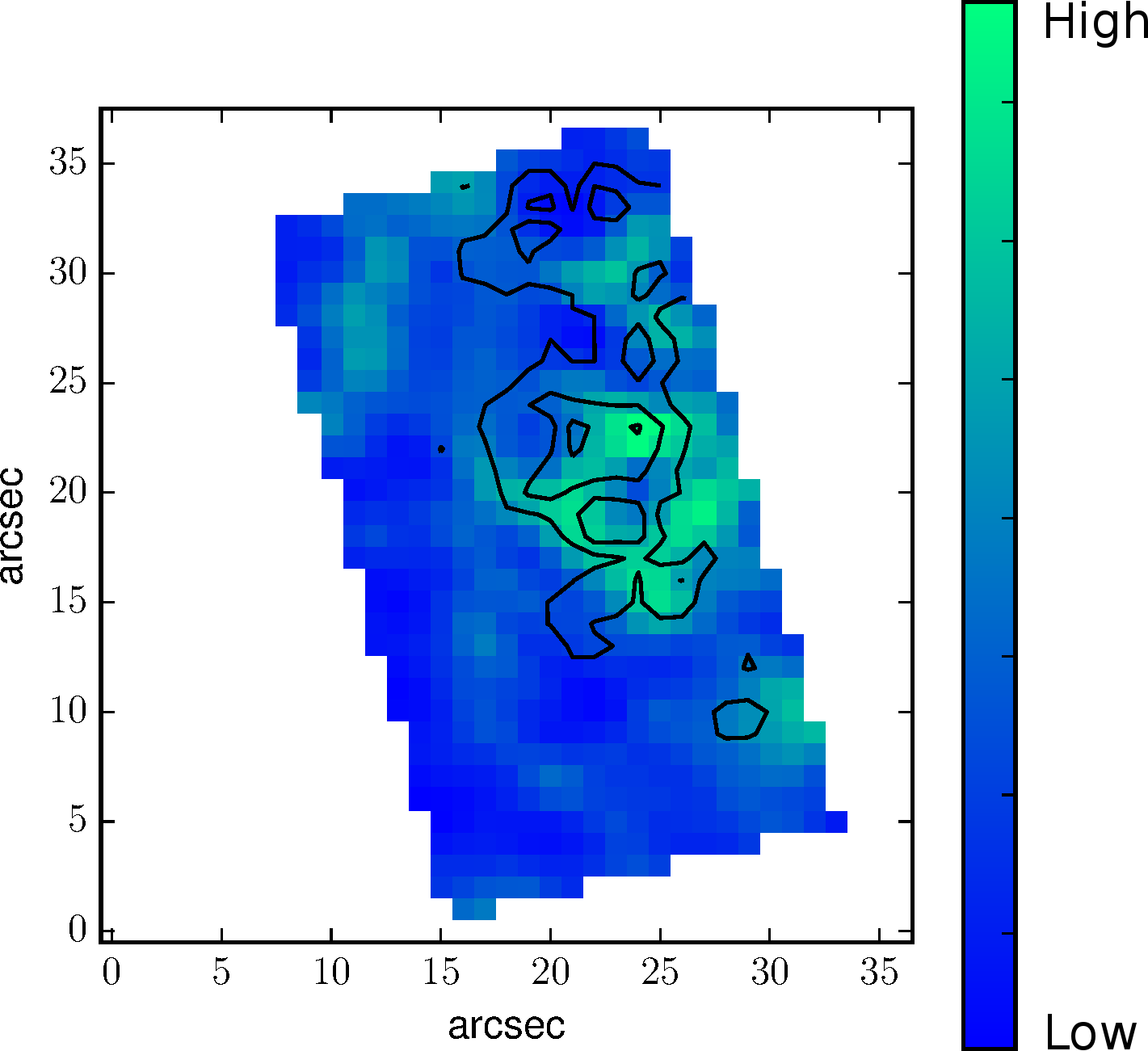}
                \includegraphics[width = 0.4 \linewidth]{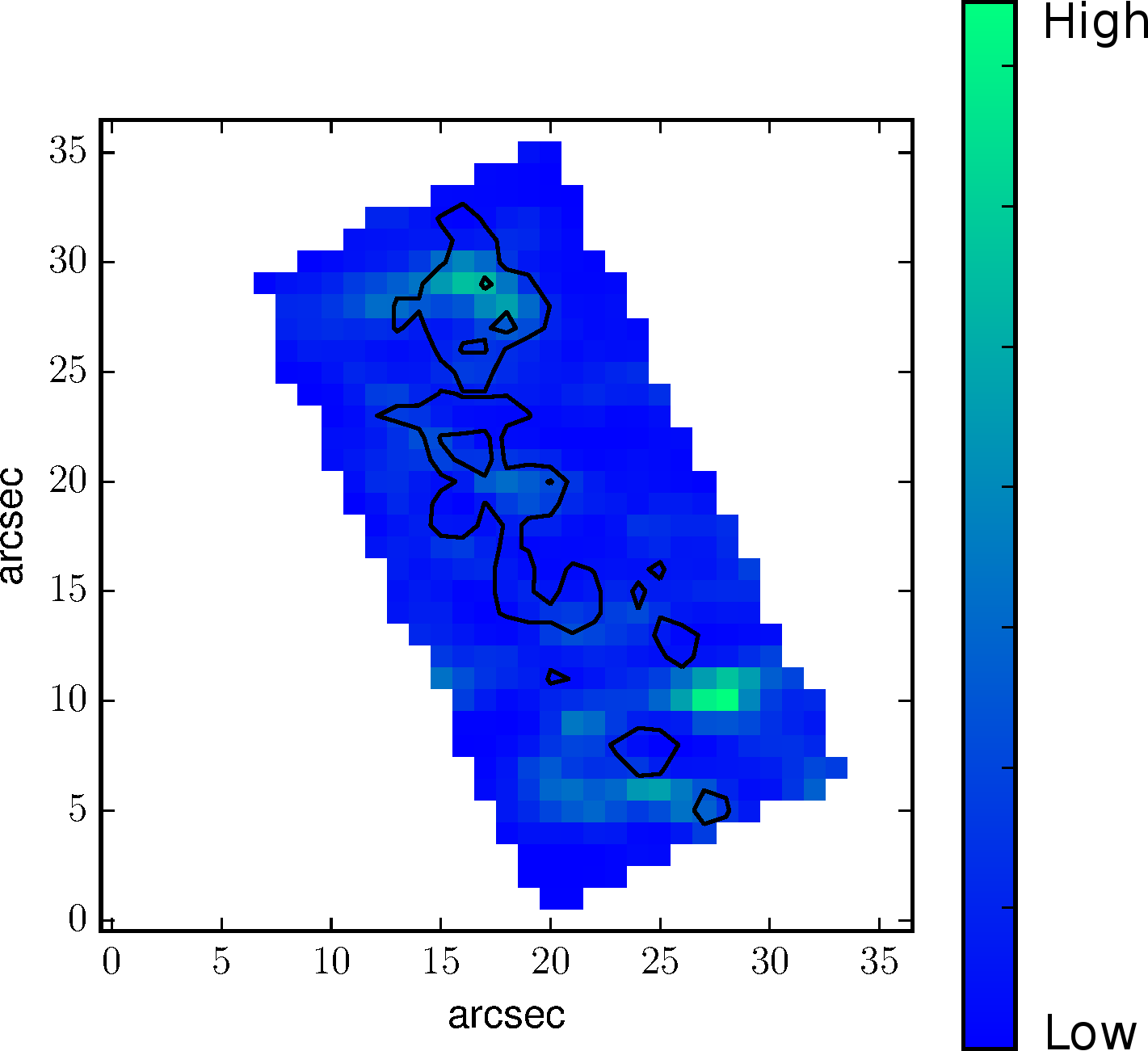}
                \caption{Maps of the pressure-based relative SFR of GalA (left) and GalB (right) and indicated by colour maps. Top: Contours of the H$\alpha$ emission at $3 \times 10^{-16} \; \si{erg \; s^{-1} \; cm^{-2}}$ and from 5 to 20$\times 10^{-16} \; \si{erg \; s^{-1} \; cm^{-2}}$ in steps of 5$ \times 10^{-16} \; \si{erg \; s^{-1} \; cm^{-2}}$ for GalA and at 5, 10, and 15$ \times 10^{-16} \; \si{erg \; s^{-1} \; cm^{-2}}$ for GalB over-plotted. Bottom: Contours of the NUV emission from GALEX over-plotted on the same pressure-based SFR colour maps as in the top panels. The optical data cubes are aligned using the DSS observations and are therefore correct to $\sim 1 - 2 \arcsec$.}
                \label{FigPressureBasedSFR}
        \end{figure*}
        
                \begin{figure*}
        \begin{center}
        \includegraphics[trim = 2 2 30 75, clip, width=0.48\linewidth]{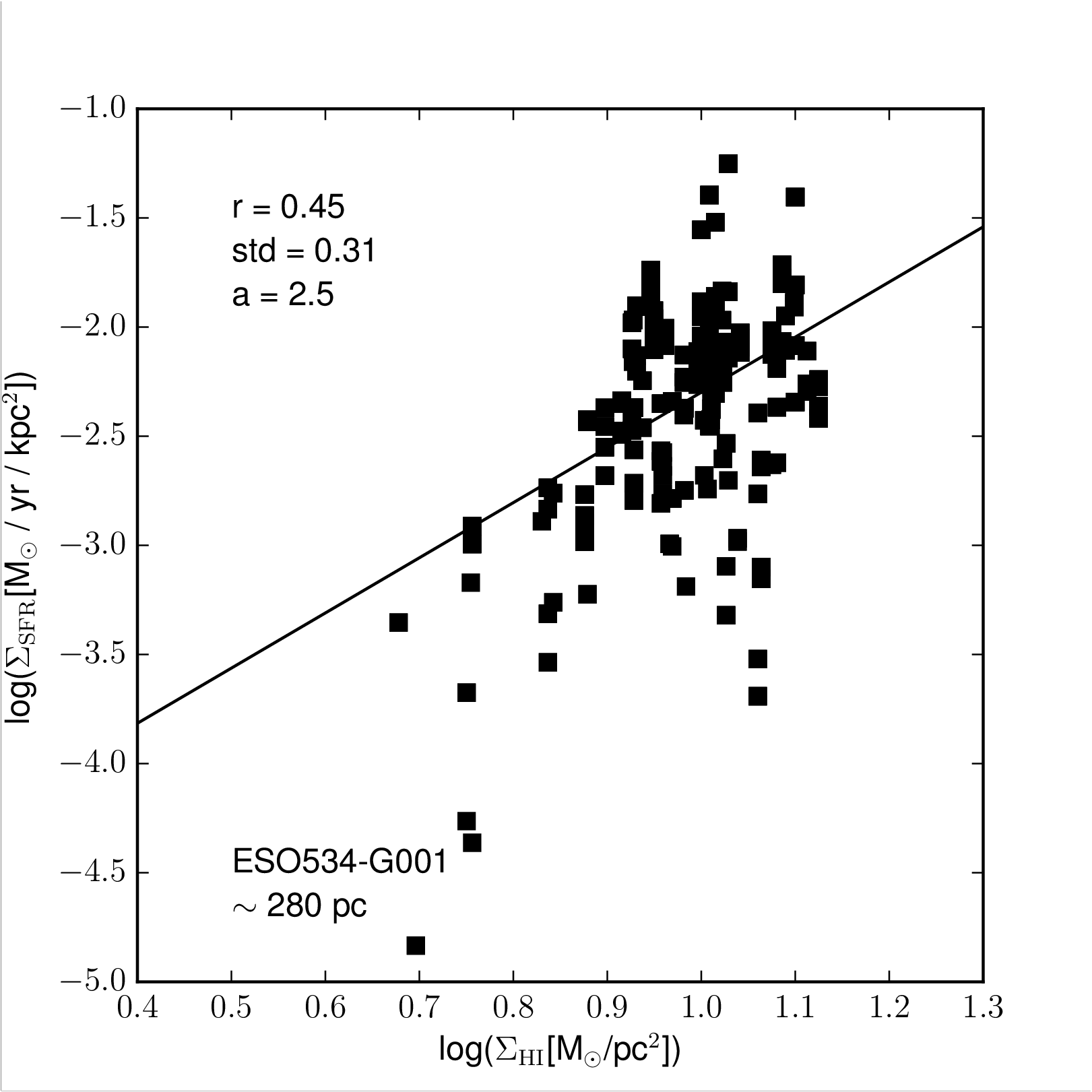}
        \includegraphics[trim = 2 2 30 75, clip, width=0.48\linewidth]{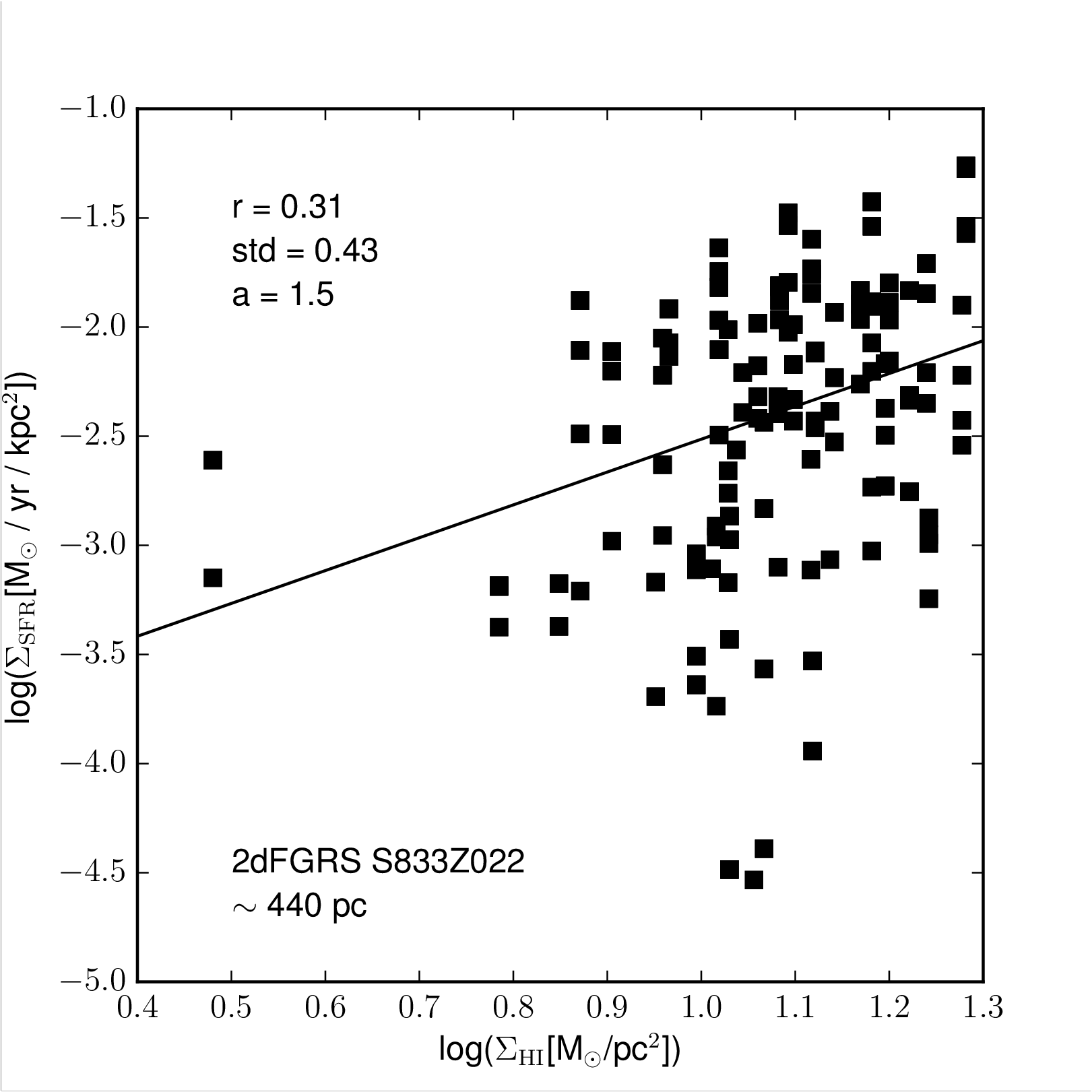}

        \includegraphics[trim = 2 2 30 75, clip, width=0.48\linewidth]{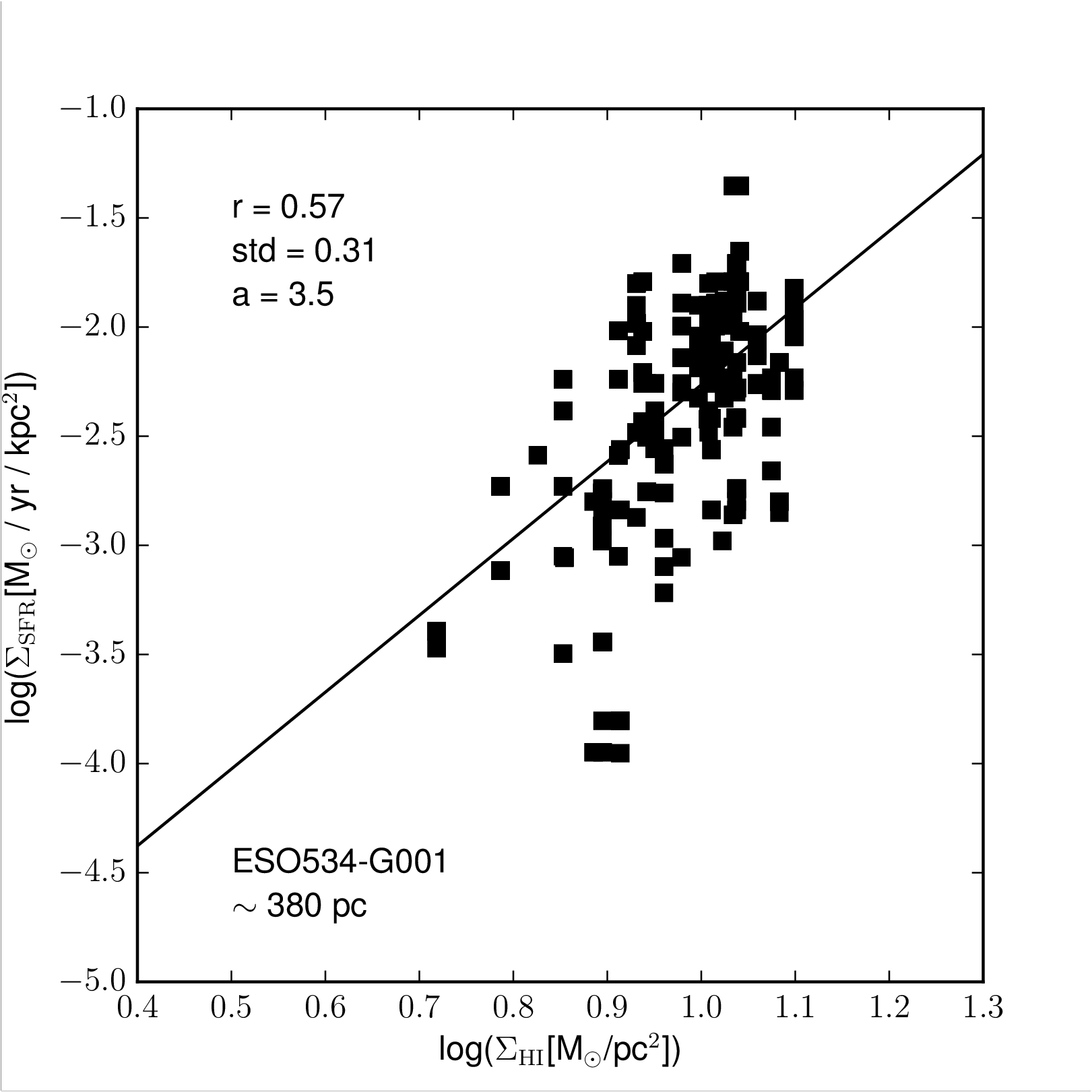}
        \includegraphics[trim = 2 2 30 75, clip, width=0.48\linewidth]{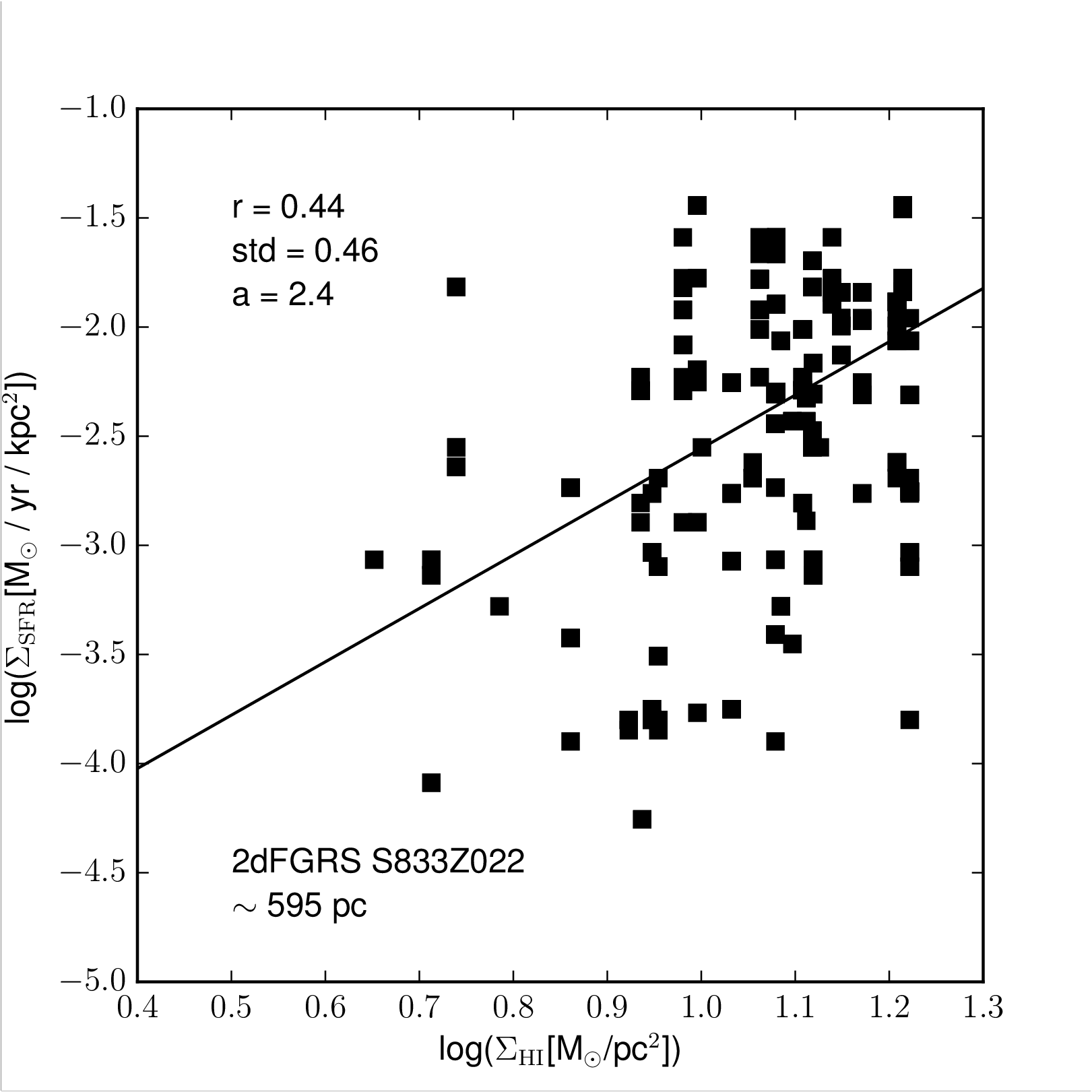}
        \end{center}
        \caption{$\Sigma_{\rm SFR}$ vs. $\Sigma_{\si{HI}}$ of GalA (left) and GalB (right). The solid line shows a power-law fit to the data. The spatial resolution for each plot is given in the lower left corner. r is the correlation coefficient, std is the corrected sample standard deviation, and a is the power-law index. From top to bottom, the spatial resolution decreases in steps of $\sim$100~pc for GalA and $\sim$150~pc for GalB. The highest spatial resolution presented is $\sim$280~pc for GalA and $\sim$440~pc for GalB.}
        \label{FigSchmidtKennicuttDecreasingRes}
\end{figure*}
\setcounter{figure}{4}
\begin{figure*}
        \center
        \includegraphics[trim = 2 2 30 75, clip, width=0.48\linewidth]{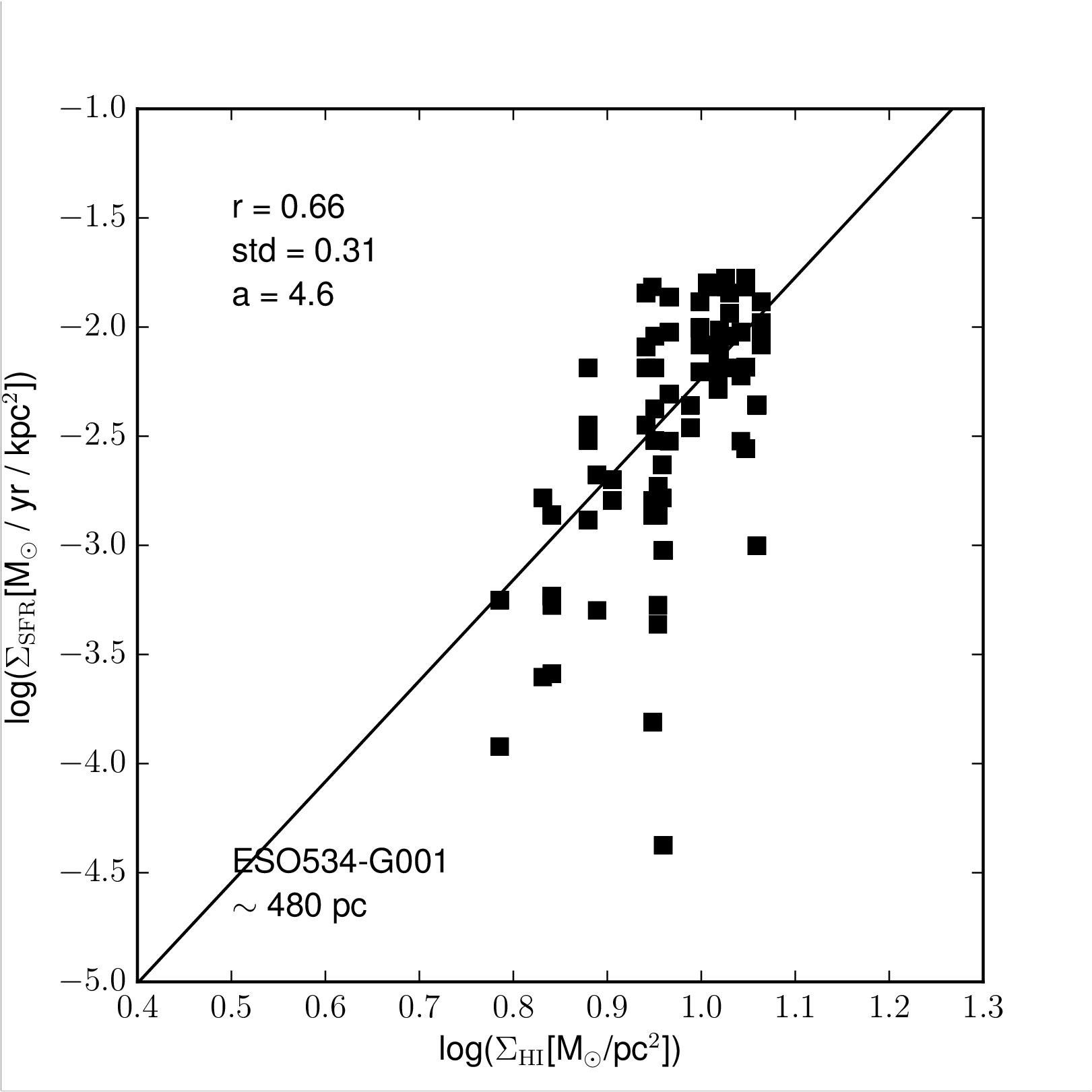}
        \includegraphics[trim = 2 2 30 75, clip, width=0.48\linewidth]{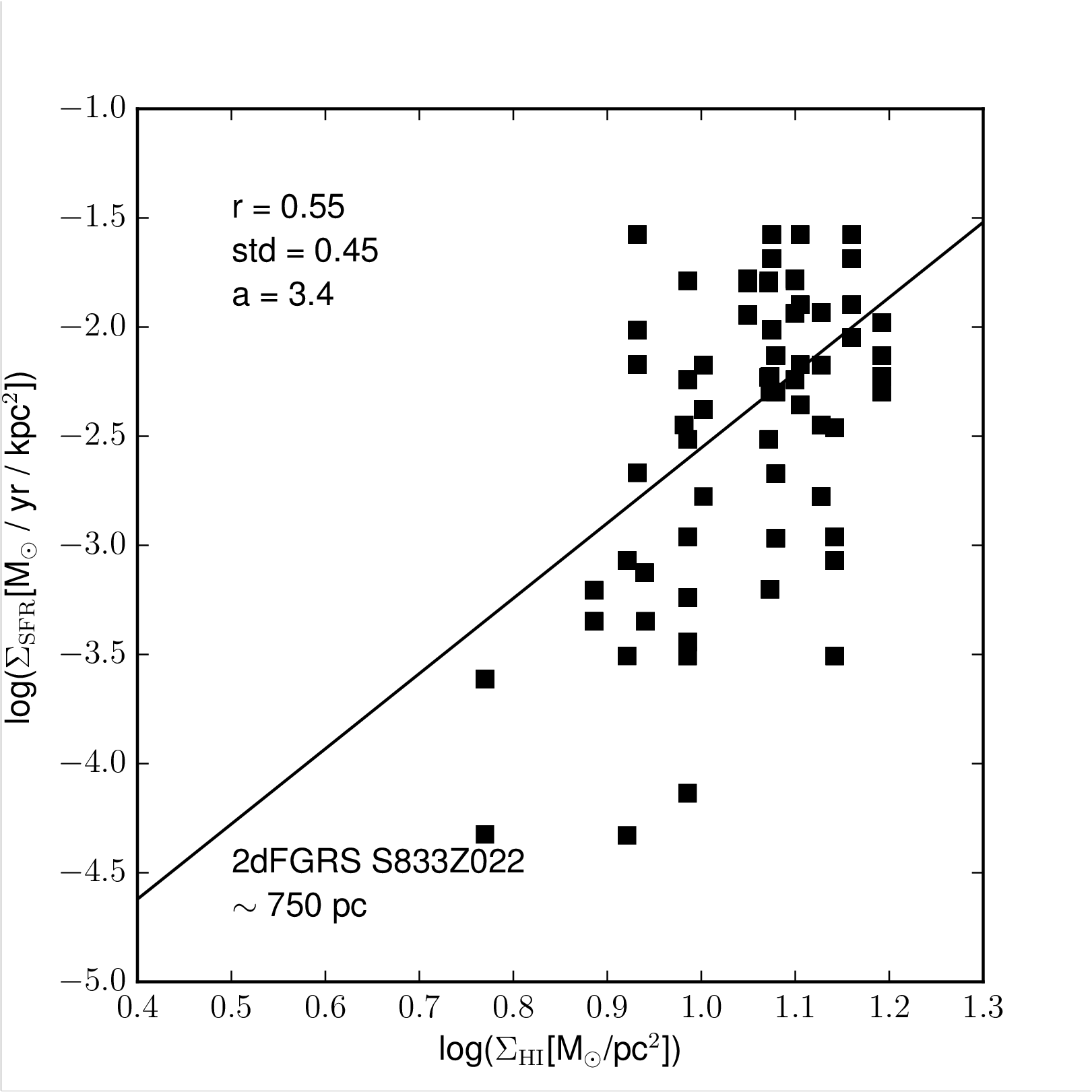}

        \includegraphics[trim = 2 2 30 75, clip, width=0.48\linewidth]{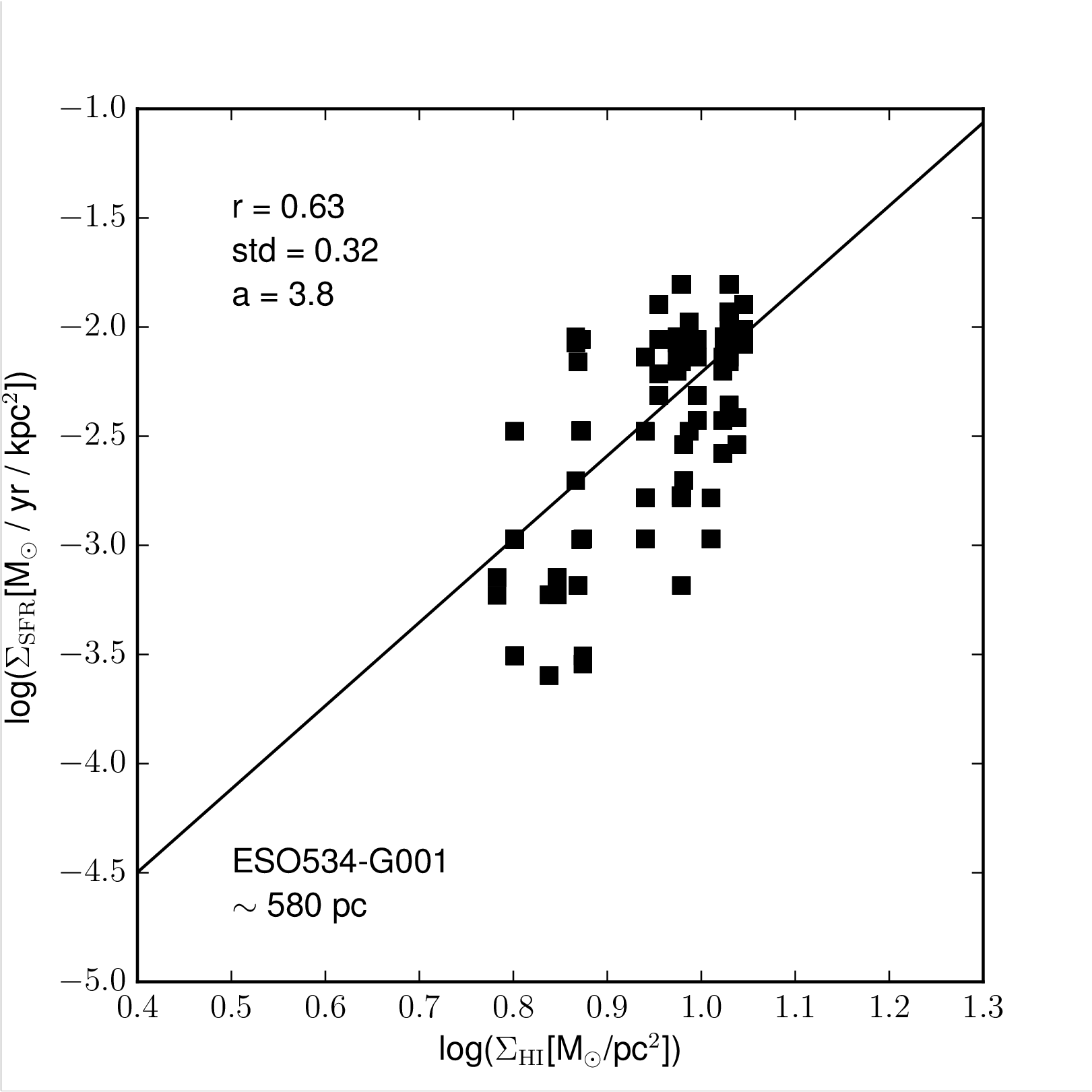}
        \includegraphics[trim = 2 2 30 75, clip, width=0.48\linewidth]{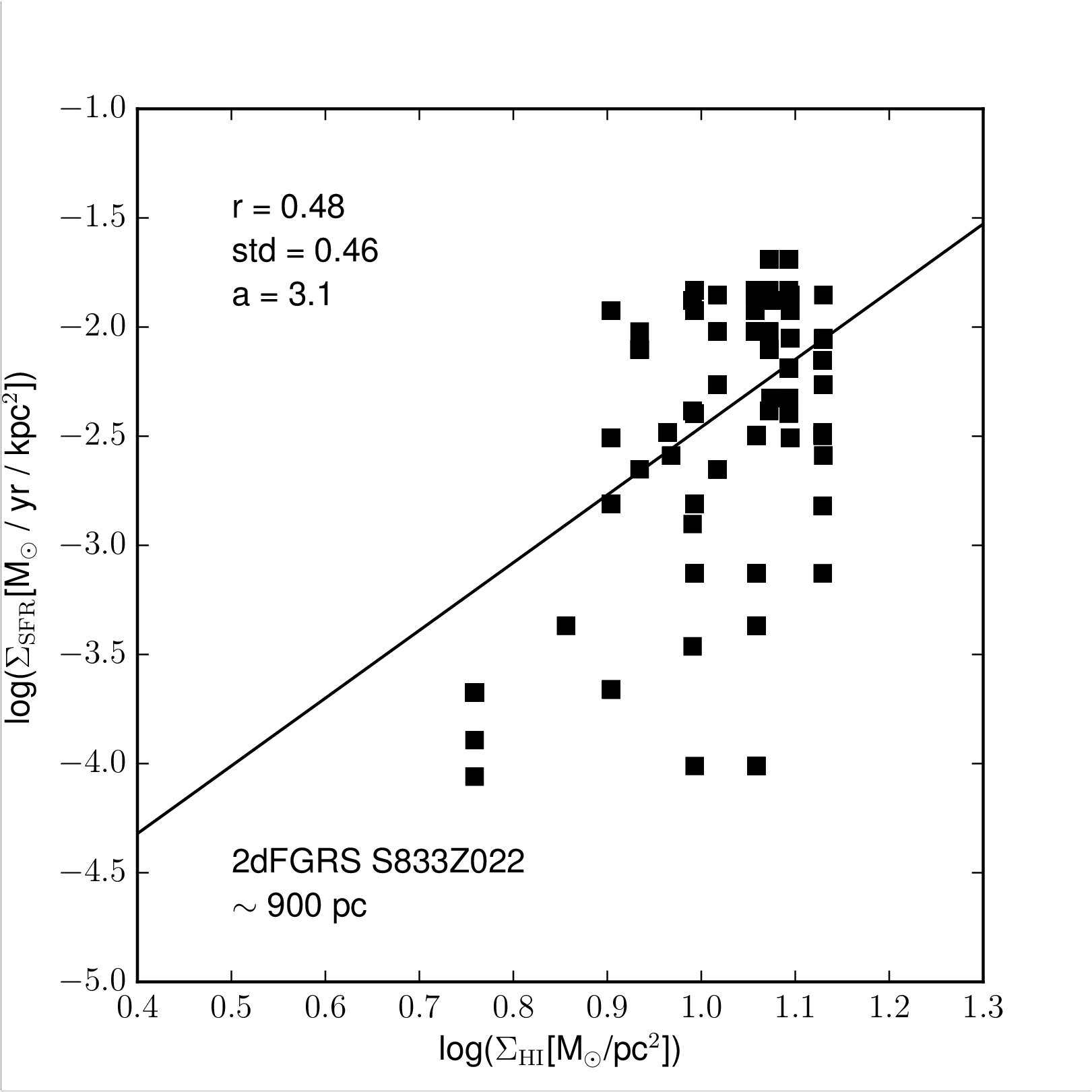}

        \includegraphics[trim = 2 2 30 75, clip, width=0.48\linewidth]{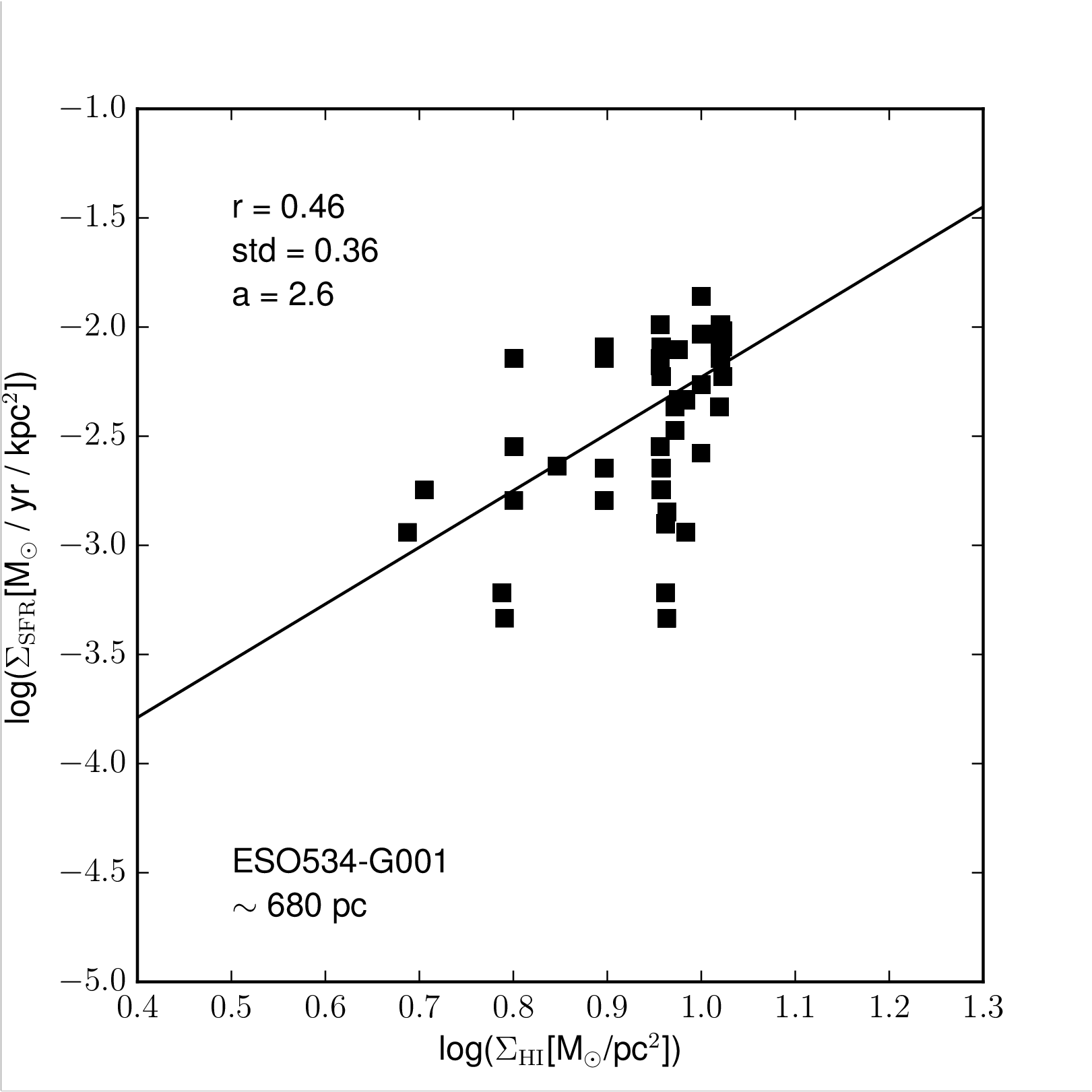}
        \includegraphics[trim = 2 2 30 75, clip, width=0.48\linewidth]{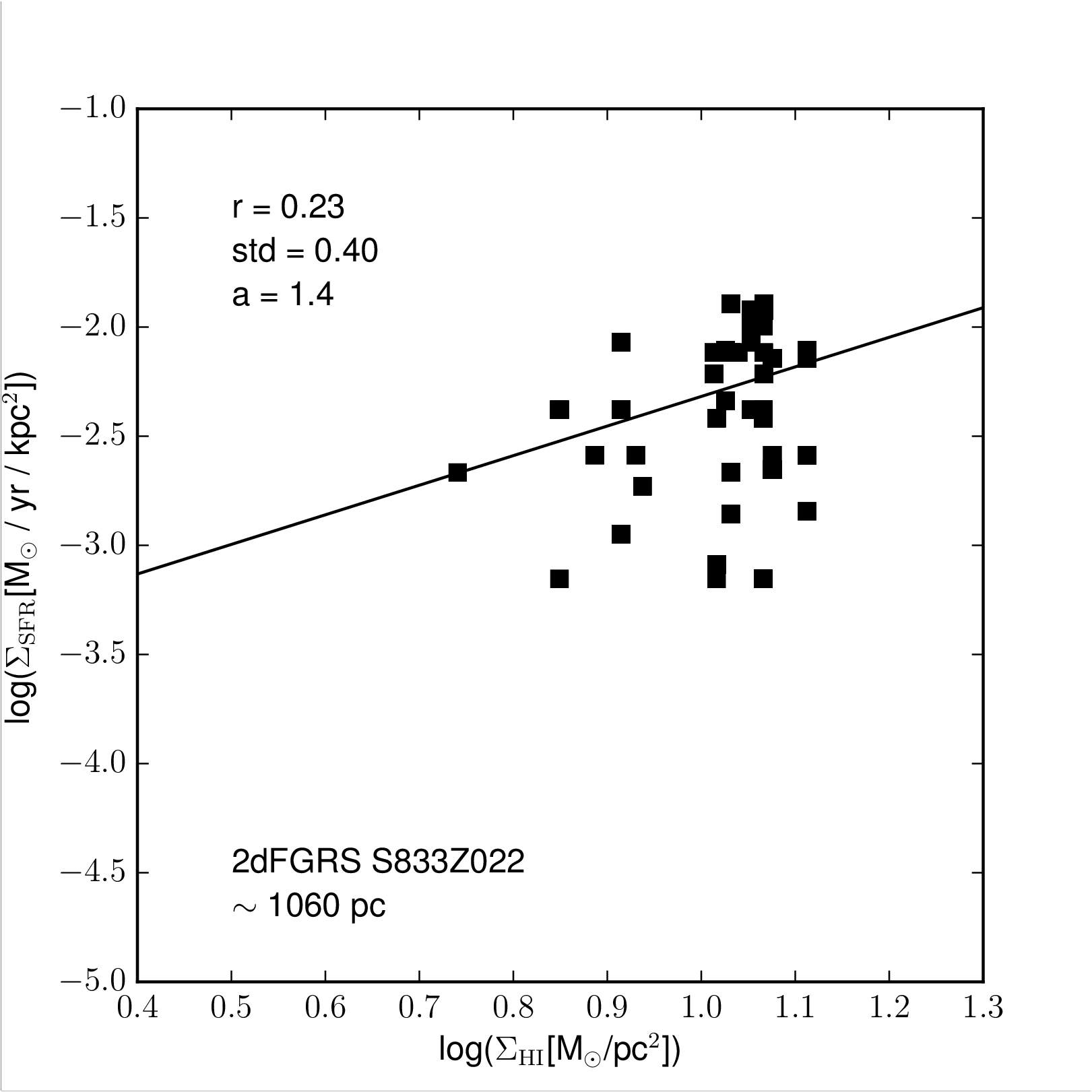}
        \caption{continued}
\end{figure*}

\begin{figure}
        \includegraphics[trim = 2 2 0 0, clip,width = \linewidth]{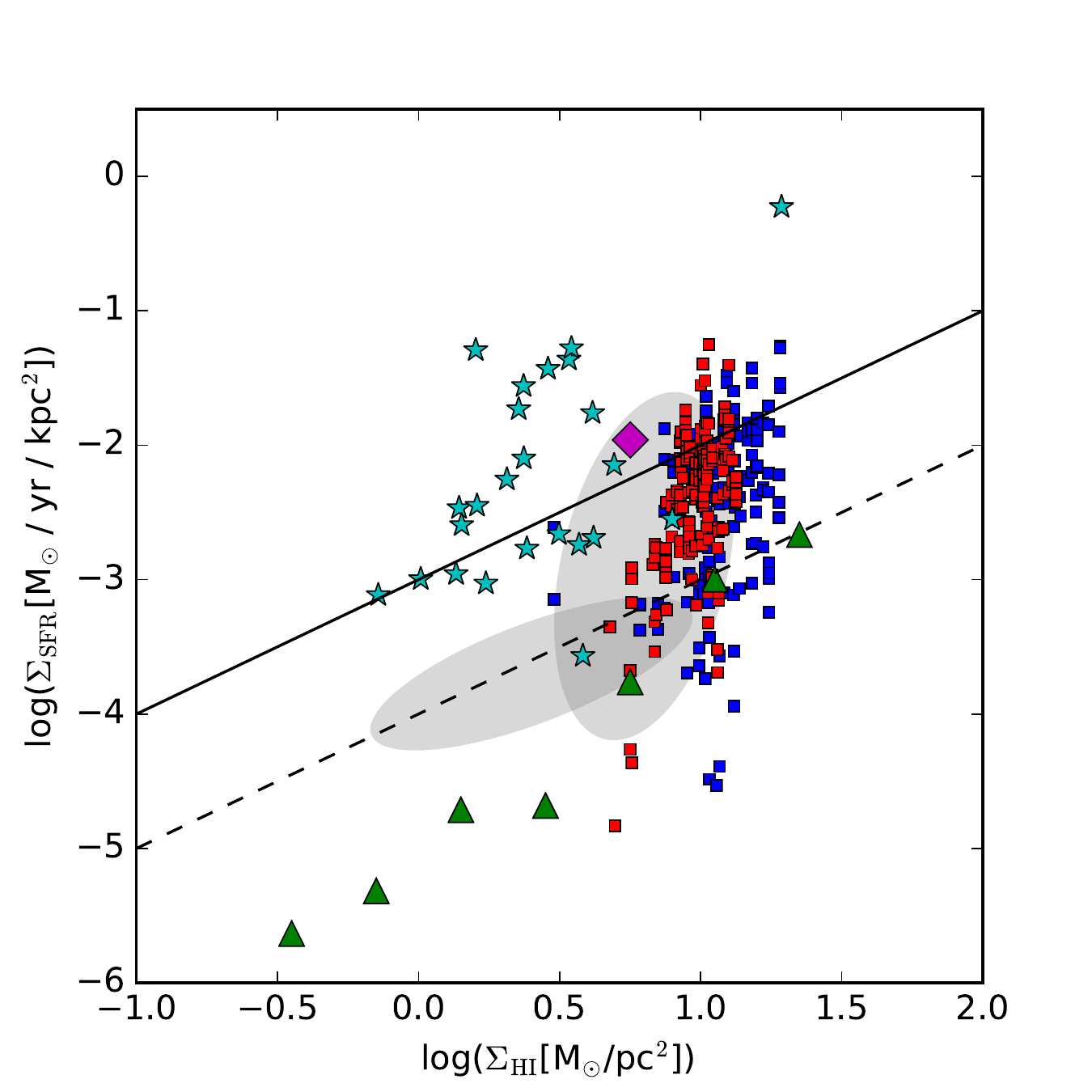}
        \caption{Schmidt-Kennicutt plot for the total
amount of gas. The green triangles are the average $\Sigma_{\rm SFR}$ within $\Sigma_{\rm HI}$ bins of 400 pc sized regions within a sample of dwarf galaxies presented by \citet{roychowdhury2015spatially}, the magenta diamond shows data from the LMC \citep{stavley-smith2003new}, the cyan stars are disc averages of local dwarf galaxies from the LITTLE THINGS Survey \citep{zhang2012hipower}, the grey shaded area represents the region in which the bulk of the spatially resolved data points from local spiral galaxies presented by \citet{bigiel2008star} at a resolution of 750 pc lie, and the blue and red squares are the spatially resolved data points of GalA and GalB at the highest resolution, respectively. The solid and dashed lines represent lines of constant atomic gas depletion times, where solid is 1 Gyr and dashed 10 Gyr.}

        \label{FigKSComp}
\end{figure}

                In this section we calculate SFRs based on the previously calculated quantities. We derive the SFR by two different approaches.
                
                The first is based on the observed H$\alpha$ emission.
                Even though these galaxies are E+A galaxies,
which should not show H$\alpha$ emission by definition, \citet{pracy2014integral} observed clumpy H$\alpha$ emission at larger radii. For the original classification this emission was not known because it was based on a single-fiber spectrum of the central region. We determine the SFR from the H$\alpha$ emission from the outer regions of the galaxies.
                \citet{murphy2011calibrating} found the following relation between H$\alpha$ luminosity $L_{\rm H\alpha}$ and SFR:
                \begin{equation}
                        \si{SFR}_{H\alpha}[\si{M}_{\odot} \; yr^{-1}] = 5.37 \times 10^{-42} L_{\rm H\alpha}[\si{erg \; s^{-1}}],\end{equation}                          
                where $L_{\rm H\alpha}$ is corrected for dust extinction. From this we obtain integrated SFRs of 0.07~M$_{\odot}$~yr$^-1$ for GalA and 0.3~M$_{\odot}$~yr$^-1$ for GalB. Based on the upper limits of the H$_2$ mass (within the region covered by the IFU observations), the corresponding gas depletion times $t_{\rm depl}$ are 18~Myr and 11~Myr. In comparison, nearby main-sequence galaxies have molecular depletion times of $\sim 2$~Gyr \citep[e.g.][]{leroy2008star, bigiel2008star}. The computation of $t_{\rm depl}$ depends on $\alpha_{\rm CO}$ , which is poorly constrained in dwarf galaxies (see above). It is therefore possible that our assumed estimate of $\alpha_{\rm CO}$ is too small. On the other hand, gas depletion times of $\sim 2$~Gyr may only hold for main-sequence galaxies in equilibrium. GalA and GalB are dwarf galaxies, and \citet{hunt2015molecular} have shown that the molecular gas depletion times in such galaxies can differ by a factor of $\sim 200$. They found a decreasing $t_{\rm depl}$ with decreasing stellar mass, consistent with other studies \citep[e.g.][]{huang2014variation}.
                
                The second approach we use for estimating the SFR is indirect and was introduced by \citet{elmegreen1989molecular, elmegreen1993h}. It is based on the atomic gas maps and assumes that the hydrostatic pressure has an impact on both the formation and destruction rate of H$_2$ and a probability of an overdensity gravitationally collapsing in a turbulent ISM. 
                We use the empirical law determined by \citet{blitz2006role} with the updated values from \citet{leroy2008star}:
                \begin{equation}
                        \frac{\Sigma_{\si{SFR}}}{\Sigma_{\si{gas}}} \propto R_{\si{mol}} \propto \left(\Sigma_{\si{gas}} \left( \Sigma_{\si{gas}} + \frac{\sigma_{\si{gas}}}{\sigma_{\star}} \Sigma_{\star} \right)\right)^{0.8},
                        \label{EQPressureBasedSFE}
                \end{equation}
                where $R_{\si{mol}} = \Sigma_{\si{H_2}}/\Sigma_{\si{HI}}$, $\Sigma_{\si{gas}}$ is the gas mass surface density, $\Sigma_{\star}$ is the stellar mass surface density, $\sigma_{\si{gas}}$ is the velocity dispersion of the gas, and $\sigma_{\star}$ is the velocity dispersion of the stars.
                We use $\Sigma_{\si{gas}}$ and $\sigma_{\si{gas}}$ maps shown in Fig. \ref{FigAnalysisResultsESO} (a) and (c) and Fig. \ref{FigAnalysisResults2dF} (a) and (c), respectively. The spatial resolution of these maps is 390 pc and 600 pc for GalA and GalB, respectively. The velocity resolution of the corresponding HI data cube is 3.3 km s$^{-1}$. Furthermore, we use $\Sigma_{\star}$ maps shown in Fig. \ref{FigAnalysisResultsESO} (f) and Fig. \ref{FigAnalysisResults2dF} (f), respectively. For $\sigma_{\star}$ we use the values determined from the spatially integrated central spectra as global values because of the low S/N. The velocity resolution of the optical spectra used to determine the velocity dispersion is 190 km s$^{-1}$ (FWHM).
                We use the HI maps as well as the IFU spectra to predict the SFR maps if Eq. (\ref{EQPressureBasedSFE}) holds. The SFR surface density predictions are shown in Fig.~\ref{FigPressureBasedSFR}. 
                The contours in these plots represent the H$\alpha$ emission (top panels) and the NUV emission, which are both linearly proportional to the SFR, but trace different timescales.
                In all plots in this figure, a mismatch between predicted and observed SFR is visible.
                The predicted SFR maps peak at different positions than the H$\alpha$  and NUV maps. The H$\alpha$  and NUV maps look similar, but are not entirely consistent, which might be caused by the two indicators probing different SF-timescales, and/or by dust extinction.
                If the SF law applies independently of the spatial averaging scale, a mismatch between the SFR prediction and the observed SFR should not occur.
                This observation shows that in HI-dominated regions the pressure-based SF law can only be used as a global SF law (see Sect. \ref{SecUncertainty}).
                
                Since the pressure-based SFR traces the HI distribution, we investigate this mismatch by a closer look at the relation between the HI distribution and H$\alpha$ flux.
                To show this, we plot in Fig. \ref{FigSchmidtKennicuttDecreasingRes} the observed $\Sigma_{\rm HI}$ versus the observed $\Sigma_{\rm SFR}$ for each resolution element.
                In addition, we fit a power law to the data. 
                To determine whether the mismatch depends on the spatial resolution and to estimate at which scale the pressure-based SF law is applicable, we show the same plots for a decreasing spatial resolution. 
                For GalA starting at a maximum spatial resolution of $\sim$280~pc, the resolution elements are increased in size in steps of $\sim$100~pc to a maximum edge length of $\sim$680~pc. 
                For GalB starting at a maximum resolution of $\sim$440~pc, the resolution elements are increased in size in steps of $\sim$150~pc to a maximum edge length of $\sim$1060~pc. 
                 The correlation coefficient slightly increases with decreasing resolution, except for the lowest resolution plots.
                The fits show that the power-law index first increases and then decreases again for a decreasing resolution. This is comparable to what was found by \citet{bigiel2008star}. 
                Furthermore, a large scatter in atomic gas depletion time is observed for the original spatial resolution. 
                 In this figure no evolution of the scatter around the respective best fit is seen when comparing the corrected sample standard deviation. To further study the deviation from a linear relation and the corresponding scatter, we apply a recently developed theoretical model in the following subsection.
                 
                 We note that based on previous studies \citep[e.g.][]{bigiel2008star}, the gas is expected to be molecular at the $\Sigma_{\rm HI}$ presented in Fig. \ref{FigSchmidtKennicuttDecreasingRes} (highest resolution). However, these studies were carried out at a lower spatial resolution. When we take the spatial resolution into account and compare the spatially resolved data points from \citet{bigiel2008star} with our data at a spatial resolution of $\sim$750 pc, the two results agree better. Furthermore, the extremely low molecular gas mass is inferred from the CO(1-0) emission line, which might
not be tracing the molecular gas well in this case.
                
                To determine whether the $\Sigma_{\rm SFR}$ we show in Fig. \ref{FigSchmidtKennicuttDecreasingRes} is atypical for local dwarf galaxies, we plot the spatially resolved data with the highest resolution onto the atomic gas Schmidt-Kennicutt relation together with a sample of spatially resolved local spiral galaxies compiled by \citet{bigiel2008star}, disc-average values from local dwarf galaxies from the LITTLE THINGS Survey \citep{zhang2012hipower}, the LMC \citep{stavley-smith2003new}, and data from a sample of spatially resolved ultra-faint dwarf galaxies presented by \citet{roychowdhury2015spatially}. The corresponding plot is shown in Fig.~\ref{FigKSComp}. Compared to the ultra-faint dwarfs, GalA and GalB have higher SFR surface densities. However, the mean stellar mass of the ultra-faint dwarf galaxies is $\sim 10^7$~M$_{\odot}$, which is two orders of magnitude lower than our galaxies. Compared to the dwarf galaxies in the same mass range from the LITTLE THINGS survey, the SFR surface densities are comparable and the LMC also has a comparable $\Sigma_{\rm SFR}$. Furthermore, when we compare our galaxies with a spatially resolved sample of local spiral galaxies from \citet{bigiel2008star}, we also find comparable $\Sigma_{\rm SFR}$. Therefore, we conclude that the $\Sigma_{\rm SFR}$ is not only comparable to disc-averaged values of dwarf galaxies with comparable mass, but also to spatially resolved data of spiral galaxies.
                Furthermore, we over-plot in Fig. \ref{FigKSComp} lines of constant gas depletion times. The atomic gas depletion time is mostly between 1 and 10 Gyr for both galaxies, which is not atypical, as can be seen by comparison to other data sets. As we have pointed out, the extremely short molecular gas depletion times result most likely from the uncertainty of the CO-to-H$_2$ conversion factor.

        \subsection{Understanding the scale-dependence of the star formation relation} \label{SecUncertainty}
        
        \begin{figure}
                \center
                \includegraphics[trim = 2 3 0 0, clip, width=\linewidth]{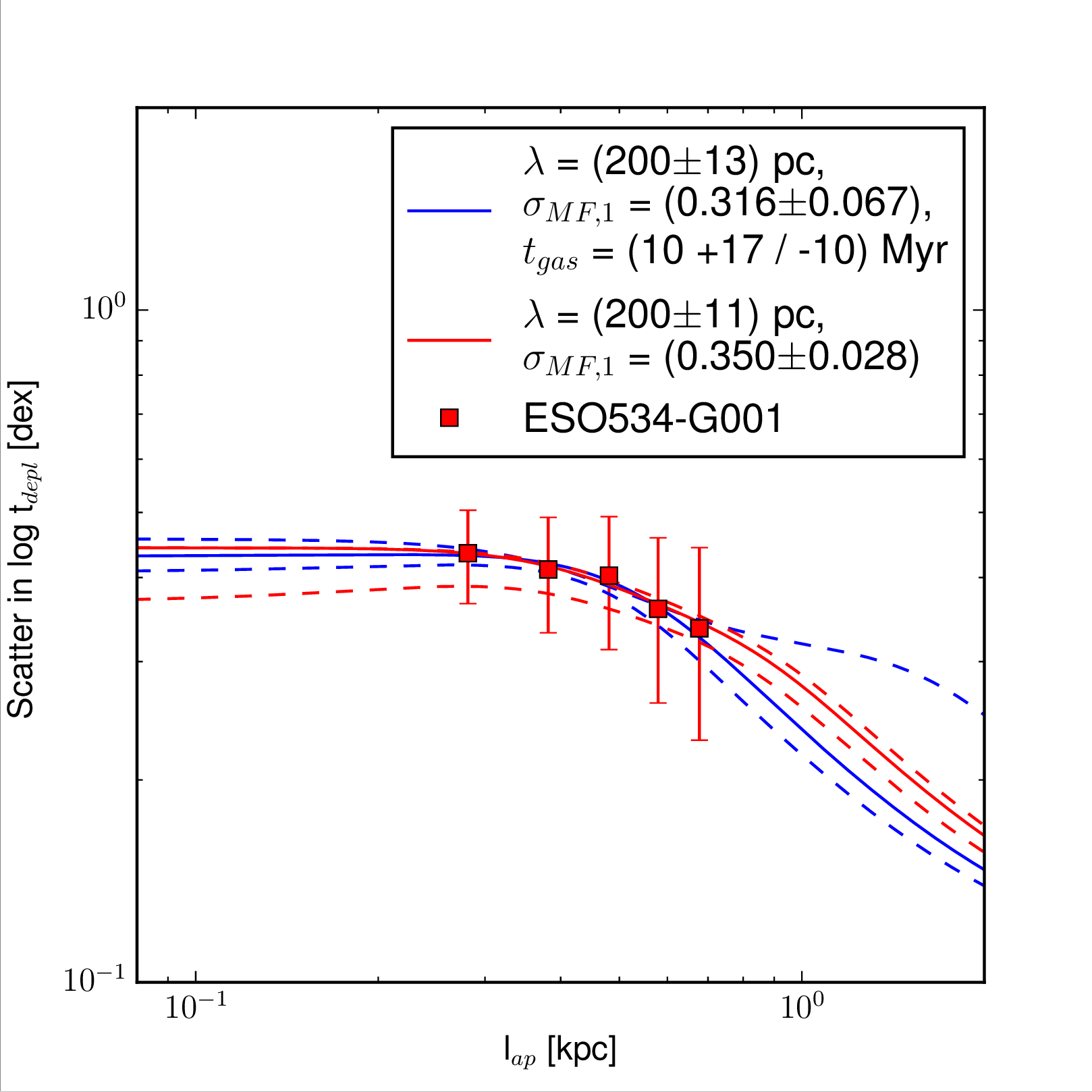}

                \includegraphics[trim = 2 2 0 0, clip, width=\linewidth]{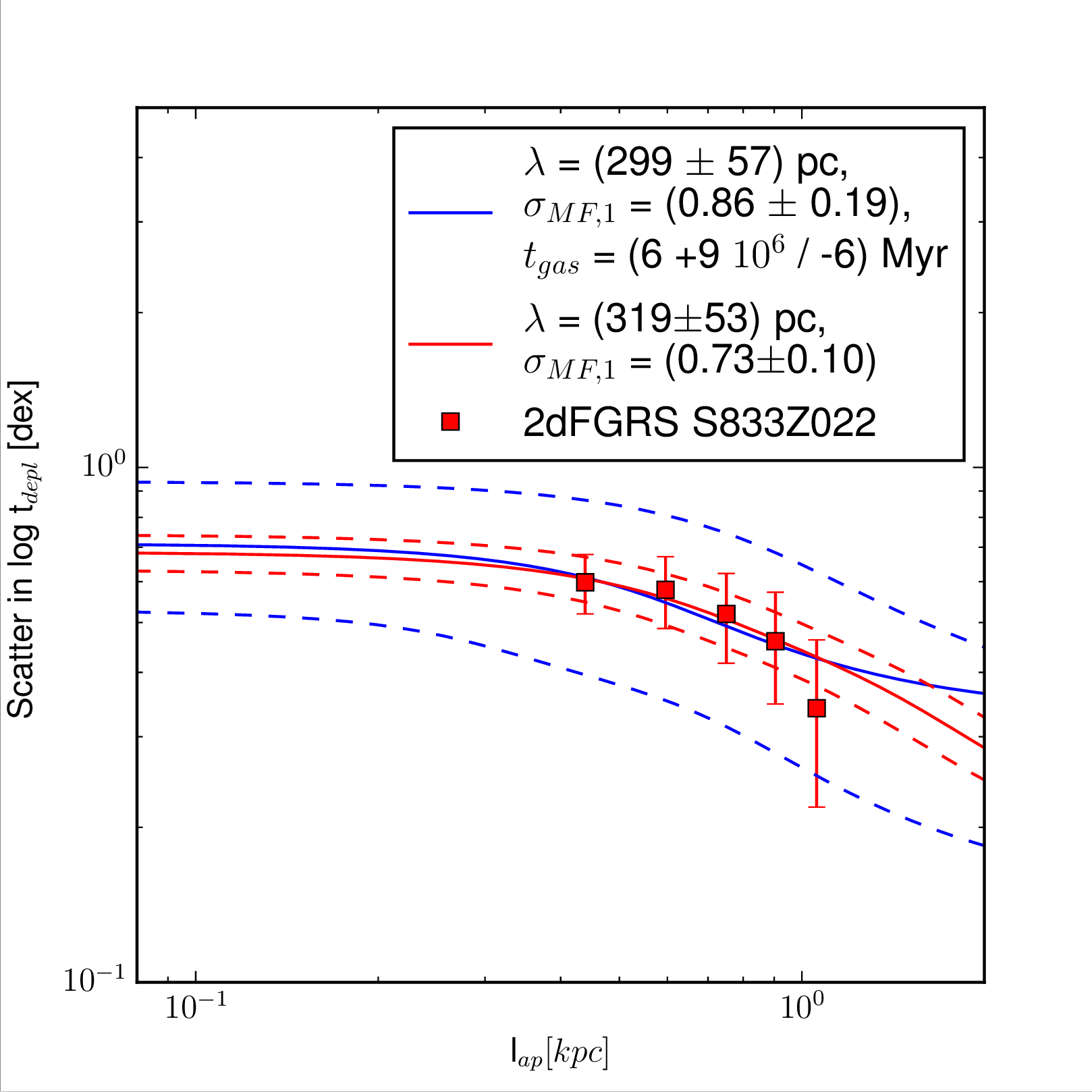}
                \caption{Scatter in gas depletion time $\sigma( \log t_{\si{depl}})$ as a function of aperture size for GalA (top) and GalB (bottom). The solid curves are least-squares fits. The red lines are two-parameter fits, and the blue solid curves are three-parameter fits, where the determined parameters and uncertainties are given in the legend. For the red curves $t_{\si{gas}}$ is kept fixed at 25~Myr. For the dashed curves the minimum/maximum parameter values of each fit are used to give an error interval.}
                \label{FigScatterTDepl}
        \end{figure}
        
We compare the scale-dependence of the SF relation (i.e.~the mismatch between the HI-predicted and observed H$\alpha$ maps) to a theoretical model for the multi-scale nature of the SF relation (the `uncertainty principle for star formation’ of \citet{kruijssen2014uncertainty}, hereafter referred to as KL principle). Galactic SF relations assume that all phases of the SF timeline are well sampled within each aperture. However, the KL principle shows that this assumption is no longer valid on small-size scales. In small apertures, the sampling of the SF process is incomplete, as it matters {\it \textup{when}} a region is observed. The corresponding prediction is that the scatter on the SF relation increases towards smaller aperture sizes. We note that the model of KL14 is entirely independent of the tracers used and only assumes that both tracers each occupy their own part of the evolutionary timeline (even though the model allows them to be completely coincident; in that case, there is no small-scale breakdown of the SF relation).

The KL principle models the scale dependence of the SF relation as a Poisson process. It assumes a random distribution of points (`regions’) in two-dimensions with a typical separation length $\lambda=2\sqrt{A/\pi N}$, where $A$ is the total area and $N$ indicates the total number of points. Each of these points represents an independent star-forming region and resides at a random moment on the evolutionary timeline. For simplicity, we assume a simple two-phase SF process, where during the first phase only the gas tracer can be observed, and during the second phase only the SF tracer can be observed, meaning that~the possible overlap time of the two phases is set to zero. While the overlap is obviously non-zero in nature, it has been shown by \citet{kruijssen2014uncertainty} that this assumption does not affect the results on the size
scales $l_{\rm ap}>\lambda$ considered here (see below).

The \citet{kruijssen2014uncertainty} model for the scatter as a function of the aperture size is described in their Appendix B and is publicly available at http://www.mpa-garching.mpg.de/KL14principle. To compare its predictions to our observations, we first derive the observed scatter in the gas depletion time ($t_{\rm depl}\equiv M_{\rm gas}/{\rm SFR}$) as a function of aperture size. To do so, we convolve the observed maps with a top-hat kernel from the original spatial resolution of $\sim$280~pc in steps of $\sim$100~pc (GalA) and $\sim$440~pc in steps of $\sim$150~pc (GalB). The final resolution is limited to 0.5 times the minor axis of the galaxies, which is $\sim$680~pc for GalA and $\sim$1060~pc for GalB. For each convolved map, we calculate the scatter according to $\sigma_{\log(t_{\rm depl})} = \sigma_{\log \left( L_{\rm HI}/L_{\rm H\alpha} \right)}$, where we use the standard deviation $\sigma$ of the quotient of the observed HI emission $L_{\rm HI}$ and the observed H$\alpha$ emission $L_{\rm H\alpha}$. This is the scatter orthogonal to a linear relation between the HI surface density and the H$\alpha$ emission. 

 Plots comparing the observed scatter $\sigma(\log(t_{\si{depl}}))$ for increasing aperture size with the model predictions are shown in Fig.~\ref{FigScatterTDepl}. 
 For the KL principle predictions, we adopt the parameters given in \citet{kruijssen2014uncertainty}. The model has four parameters, of which one is fixed in advance: $t_{\rm gas}$ is the duration of the gas phase, $t_{\rm star}$ is the duration of the young stellar phase, $\lambda$ is the separation length of independent regions, and $\sigma_{\rm MF1}$ is the scatter induced by the fact that the mass spectrum of star-forming regions is not a delta function. Of these parameters, we fix $t_{\rm star}$ to a value of 6~Myr because we use H$\alpha$ to trace the SF \citep[cf.][]{leroy2013molecular}. The other three parameters are left free and specified in the legend of Fig. \ref{FigScatterTDepl}. As shown in Fig.~3 of KL14, the size scale of the knee in the curves in Fig.~\ref{FigScatterTDepl} is typically $\sim2.5$ times larger than $\lambda$. As a result, it is possible to constrain the separation length between independent cloud and star-forming regions using Fig.~\ref{FigScatterTDepl}. We attempt to constrain the free parameters by performing a least-squares fit of the KL14 model to the data, where the weights are the inverse errors of the observed scatter. We perform both a two- and a three-parameter fit, where $t_{\si{gas}}$ is either fixed to 25~Myr or left free. We find that $t_{\si{gas}}$  only has an impact on the predicted scatter at large aperture sizes. Because of the small angular size of the observed galaxies, we cannot trace size scales large enough to constrain $t_{\rm gas}$. On the other hand, $\lambda$ and $\sigma_{\si{MF1}}$ can both be determined quite accurately from the fits, although we caution against overinterpreting these numbers as the parameters are strongly correlated, with correlation coefficients larger than 0.7. This confirms the finding of \citet{kruijssen2014uncertainty} that the scatter as a function of size scale is a degenerate result of (most of) the underlying parameters that describe the cloud-scale SF process. The authors showed that these parameters can be constrained to very high accuracy by instead considering the bias of the depletion time when focusing apertures on gas peaks or young stellar peaks. Even though such an analysis could in principle be carried out for the two galaxies considered here, it would require a larger number of peaks than can be identified at the current resolution of our maps \citep{kruijssen16}.
                
\section{Summary and conclusion}

        We presented a multi-tracer study of the two galaxies ESO534-G001 and 2dFGRS~S833Z022, covering atomic, molecular, and H$\alpha$ emission. These galaxies are known to have an E+A-like optical spectrum in the centre and SF at larger radii observed in optical IFU data as well as in GALEX NUV emission images. 
        The goal was to study the SF relation in these systems, in which SF might proceed differently than in main-sequence galaxies. 
        The results of our analysis are as follows.

                The observations of the HI 21~cm emission line show that the neutral hydrogen gas is distributed as expected for typical spiral galaxies throughout the entire discs of both galaxies. 
                In the velocity maps we observe a regularly rotating disc.
                The velocity dispersion maps reveal no peculiarities. The observed velocities and dispersions are comparable with regular star-forming spiral galaxies.
                
                We did not detect the CO(1-0) emission line. 
                This implies either that the H$_2$ mass is low, or, alternatively, that the H$_2$ is not well traced by CO in these galaxies. From the detection limit, it is only possible to calculate upper limits of 1.3~$\times10^6$~M$_{\sun}$ (ESO534-G001) and $3.3 \times 10^6$~M$_{\sun}$ (2dFGRS~S833Z022). These upper limits are only valid for molecular hydrogen structures on angular scales smaller than $\sim$20\arcsec or $\sim$2~kpc. To confirm the low molecular hydrogen mass, single-dish observations of CO or other dense gas tracers have to be obtained.
                
                We found short molecular gas depletion times of $\lesssim 20$ Myr in these galaxies with an E+A galaxy-like spectrum in the center. This result is strongly influenced by the uncertainty of the CO-to-H$_2$ conversion factor. Furthermore,
we propose an additional explanation. These gas depletion times are in the lower range of the observed scatter in low-mass galaxies, which reaches $<$50 Myr \citep{hunt2015molecular} and is possibly due to the bursty nature of SF in low-mass galaxies. 
                
                We used the HI maps to calculate the predicted SFR following the pressure-based law of \citet{blitz2006role}. We compared this to the observed SFR based on the H$\alpha$ and NUV emission. In these maps we found a clear mismatch between predictions and observations, in that the predicted and observed SF peaks are at different positions. This indicates that the pressure-based SF relation in HI-dominated regions only applies globally and does not describe SF on the cloud scale.
                
                We studied the small-scale breakdown of the atomic SF relation further by deriving it for different size scales. We found that the scatter around a power-law fit decreases with decreasing resolution for both galaxies. Furthermore, the power-law index increases towards the lowest resolution. These findings indicate that galactic SF relations do not apply on sub-galactic scales.
                
                We compared the observed scatter in gas depletion time with the predictions from the `uncertainty principle for star formation' of \citet{kruijssen2014uncertainty} and found that our observations agree with the predictions. This shows that the scale dependence of the {\it \textup{atomic}} SF relation in these galaxies is consistent with the predicted increase of the scatter towards small size scales that is due to the incomplete sampling of independent HI clouds and star-forming regions. This finding adds to the existing literature reporting a scale dependence of the {\it \textup{molecular}} SF relation, showing that the atomic and molecular phases are both susceptible to the evolutionary cycling of individual regions. This suggests that the atomic gas reservoirs host substantial substructure, which should be observable with future high-resolution observations.

        We found short molecular $t_{\rm depl}$, which could suggest that the two galaxies are in a transitioning phase in which SF may proceed at a different rate than in main-sequence galaxies. However, the observed total SFRs are not atypical, and the previously observed E+A-like spectrum is only a central feature. Furthermore, the galaxies lie in the blue cloud regime of the NUV-r vs. stellar mass diagram. Owing to the limited flux recovery of interferometric observations, our depletion time measurements should be confirmed with single-dish observations. Despite the non-detection of CO, we have found a substantial reservoir of atomic hydrogen in these galaxies, suggesting that the central region may be rejuvenated and experience multiple presumably bursty SF cycles before being quenched.

\section*{Acknowledgments}
The authors would like to thank Diederik Kruijssen for assistance with the application of the 'uncertainty principle for star formation' and further comments that greatly improved the paper. 
We thank Andreas Schruba for very interesting discussions about molecular gas depletion times.
We thank Achim Weiss for suggestions and discussions during this project.
We would like to thank the anonymous referee for insightful comments that helped to improve the paper.\\
The National Radio Astronomy Observatory is a facility of the National Science Foundation operated under cooperative agreement by Associated Universities, Inc.\\
This paper makes use of the following ALMA data: ADS/JAO.ALMA\# 2012.1.00293.S ALMA is a partnership of ESO (representing its member states), NSF (USA) and NINS (Japan), together with NRC (Canada) and NSC and ASIAA (Taiwan), in cooperation with the Republic of Chile. The Joint ALMA Observatory is operated by ESO, AUI/NRAO and NAOJ.\\

\end{document}